\documentclass[
 reprint,
superscriptaddress,
 amsmath,amssymb,
 aps,
]{revtex4-2}

\usepackage{graphicx}
\usepackage{dcolumn}
\usepackage{xcolor}
\usepackage{bm}
\usepackage{siunitx}
\usepackage{hyperref}

\begin{document}

\preprint{APS/123-QED}

\title{Bounds on heavy axions with an X-ray free electron laser}

\author{Jack W. D. Halliday}
 \address{Department of Physics, University of Oxford, Parks Road, OX1 3PU, UK}
 \address{Blackett Laboratory, Imperial College London,  SW7 2AZ, UK}
 \address{STFC, Rutherford Appleton Laboratory, Didcot OX11 0QX, UK}
\author{Giacomo Marocco}%
\address{Lawrence Berkeley National Laboratory,
1 Cyclotron Road, Berkeley, CA 94720-8153, USA}
\author{Konstantin A. Beyer}
\address{Max-Planck-Institut f\"ur Kernphysik Saupfercheckweg 1,
69117, Heidelberg, Germany}
\author{Charles Heaton}
\address{Department of Physics, University of Oxford, Parks Road, OX1 3PU, UK}
\author{Motoaki Nakatsutsumi}
\address{European XFEL, Holzkoppel 4, 22869 Schenefeld, Germany}
\author{Thomas R. Preston}
\address{European XFEL, Holzkoppel 4, 22869 Schenefeld, Germany}
\author{Charles D. Arrowsmith}
\address{Department of Physics,  University of Oxford, Parks Road, OX1 3PU, UK}
\author{Carsten Baehtz}
\address{Helmholtz-Zentrum Dresden-Rossendorf, Bautzner Landstraße 400, 01328 Dresden, Germany}
\author{Sebastian Goede}
\address{European XFEL, Holzkoppel 4, 22869 Schenefeld, Germany}
\author{Oliver Humphries}
\address{European XFEL, Holzkoppel 4, 22869 Schenefeld, Germany}
\author{Alejandro Laso Garcia}
\address{Helmholtz-Zentrum Dresden-Rossendorf, Bautzner Landstraße 400, 01328 Dresden, Germany}
\author{Richard Plackett} 
\address{Department of Physics, University of Oxford, Parks Road, OX1 3PU, UK}
\author{Pontus Svensson}
\address{Department of Physics, University of Oxford, Parks Road, OX1 3PU, UK}
\author{Georgios Vacalis}
\address{Department of Physics, University of Oxford, Parks Road, OX1 3PU, UK}
\author{Justin Wark}
\address{Department of Physics, University of Oxford, Parks Road,  OX1 3PU, UK}
\author{Daniel Wood}
\address{Department of Physics, University of Oxford, Parks Road, OX1 3PU, UK}
\author{Ulf Zastrau}
\address{European XFEL, Holzkoppel 4, 22869 Schenefeld, Germany}
\author{Robert Bingham}
\address{STFC, Rutherford Appleton Laboratory, Didcot OX11 0QX, UK
}
\address{John Anderson Building, University of Strathclyde, G4 0NG, UK
}
\author{Ian Shipsey}
\address{Department of Physics, University of Oxford, Parks Road, OX1 3PU, UK}
\author{Subir Sarkar}
\address{Department of Physics,  University of Oxford, Parks Road, OX1 3PU, UK}
\author{Gianluca Gregori}
\address{Department of Physics, University of Oxford, Parks Road,  OX1 3PU, UK}
\date{\today}

\begin{abstract}
We present new exclusion bounds obtained at the European X-ray Free Electron Laser facility (EuXFEL) on axion-like particles (ALPs) in the mass range $10^{-3}\;\si{\electronvolt} \lesssim m_a \lesssim  10^{4}\;\si{\electronvolt}$. Our experiment exploits the Primakoff effect via which photons can,  in the presence of a strong external electric field, decay into axions, which then convert back into photons after passing through an opaque wall.
While similar searches have been performed previously at a 3$^{\rm rd}$ generation synchrotron \cite{Yamaji:2018ufo}, our work demonstrates improved sensitivity, exploiting the higher brightness of X-rays at EuXFEL.
\end{abstract}

\maketitle
\emph{Introduction.} The axion arises from the breaking of Peccei-Quinn (PQ) symmetry \cite{Peccei:1977hh,Weinberg:1977ma,Wilczek:1977pj}, which was proposed to explain the absence of $CP$-violation by the strong interactions described by quantum chromodynamics (QCD). Axion-like particles (ALPs) also arise in string theory \cite{Svrcek:2006yi}. In spite of being very light and having suppressed couplings, coherent oscillations of relic axions can naturally account for cold dark matter if $m_a \sim 10^{-6}$--$10^{-4}\;\si{\electronvolt}$ \cite{Preskill:1982cy,Abbott:1982af,Dine:1982ah}. 
Most laboratory searches for axions converting to photons in a magnetic field \cite{Sikivie:1983ip} 
have therefore focussed on this `light axion window' \cite{Semertzidis:2021rxs}, targeting axion-photon couplings corresponding to the Galactic halo dark matter being made of axions. This coupling is related (inversely) to the scale of PQ symmetry breaking in extensions of the Standard Model that implement the PQ symmetry, e.g. the Kim-Shifman-Vainshtein-Zakarov (KSVZ) model \cite{Kim:1979if,Dine:1981rt} or the Dine-Fischler-Srednicki-Zhitnitsky (DFSZ) model \cite{Dine:1981rt,Zhitnitsky:1980tq}. It has been noted that when the PQ symmetry (in the DFSZ model) is broken after cosmological inflation, axions are also produced by the decay of domain walls \cite{Ringwald:2015dsf}, and the preferred mass for axions to make up dark matter then exceeds $10^{-2}\;\si{\electronvolt}$ \cite{Beyer:2022ywc}. Such ``heavy" axions are associated with a low  scale of Peccei-Quinn symmetry breaking, so are theoretically preferred as being less susceptible to the `axion quality problem', namely the potential destabilising effects of quantum gravity on global symmetries \cite{Kamionkowski:1992mf, Holman:1992us, Barr:1992qq}.

Stringent bounds on such heavy axions (excluding astrophysical arguments derived from stellar cooling \cite{Caputo:2024oqc}) come from the CERN Axion Solar Telescope (CAST) \cite{CAST:2017uph}. This is a `helioscope' which looks for conversion of axions from the Sun into X-ray photons as they pass through a strong magnetic field. However due to the specific experimental geometry of CAST, the axion-photon conversion probability gets highly suppressed for $m_a \gtrsim \SI{1}{\electronvolt}$. For such masses, more competitive bounds arise from experiments which exploit Bragg conversion in the electric field of crystals, and underground searches for dark matter and $\beta\beta$-decay have been claimed to place strong bounds on the axion-photon coupling \cite{SOLAX:1997lpz, Bernabei:2001ny,COSME:2001jci, CDMS:2009fba, Belli:2012zz,Armengaud:2013rta, Majorana:2022bse,XENON:2022ltv}. However when the damping of X-rays in a crystal is taken into account, such bounds are considerably weakened \cite{Dent:2023gzl}. Moreover, since the axions originate from the Sun there is necessarily some model dependence in extracting such bounds; the high plasma frequency and temperature in the Sun are particularly relevant as these can perturb the effective axion-photon coupling \cite{Jaeckel:2006xm,Caputo:2024oqc}. Similarly bounds derived from stellar cooling arguments, e.g. neutrino observations of Supernova 1987a, have large astrophysical uncertainties \cite{Bar:2019ifz}. 

By contrast in laboratory experiments the axion production process is directly controlled, avoiding such model dependence. Interesting constraints have been set by accelerator experiments, such as NOMAD \cite{NOMAD:2000usb}, BaBar \cite{BaBar:2017tiz,Dolan:2017osp}, and NA64 \cite{NA64:2020qwq}. Laboratory-based searches for axions are thus well motivated even though they do not presently reach the same sensitivity as astrophysical limits. Of course it is important to use as many different experimental approaches as is feasible, since each has its own characteristic strengths and limitations.

Here we present results from a new laboratory search for axions performed with the HED/HiBEF instrument at the EuXFEL in Hamburg \cite{Zastrau2021}. This is sensitive to a broad range of axion/ALP mass, between $\sim 10^{-3} - 10^{4}\;\si{\electronvolt}$. 
Our experiment exploits the Primakoff effect via which photons can decay into axions in the presence of a strong external electric field and then reconvert back into photons after passing through an opaque wall. This technique was previously employed in experiments with optical lasers and external magnetic fields \cite{Robilliard:2007bq,Ehret:2010mh,OSQAR:2015qdv}.

When using X-rays, it is possible to increase the detection sensitivity by exploiting the electric fields which are present within a crystalline material. These atomic electric fields can be as high as $10^{11} \; \si{\volt\per\meter}$ which, due to the form of the Hamiltonian, corresponds to magnetic field strengths of order \SI{1}{\kilo\tesla} -- much higher than the field strengths accessible using the best electromagnets. Although the length scales which are accessible with crystal based searches are smaller than for those with electromagnets, the path integrated equivalent field is competitive, being $\sim \SI{25}{\tesla\meter}$ for the present study.

The strength of the effective magnetic field is calculated numerically using a Draic-Fock method \cite{Doyle1968, Baier1998}; these calculations are well verified experimentally, for example in positron channelling experiments \cite{Atkinson1982}.      

Furthermore, arranging atoms in a crystalline structure leads to a coherent effect analogous to Bragg scattering. Generation and reconversion can thus be carried out with a pair of X-ray crystals. This concept was first described by Buchm\"uller \& Hoogeveen \cite{Buchmuller:1989rb}.

We improve on previous laboratory-based searches in the above mass range (up to 
which were performed using 3$^{\rm rd}$ generation synchrotron facilities \cite{Yamaji:2018ufo,Yamaji:2017pep} 
but we achieve higher detection sensitivity due to the increased brightness of Free Electron Lasers (FELs). 
This is because of the much shorter duration of the photon pulse which allows for a more accurate discrimination of the signal against the background.


{\it Experimental setup.} As discussed, a number of experiments have already placed bounds on the available axion parameter space, with varying degrees of model dependence. We use the term axion to describe both the QCD axion and any ALP which couples to photons via the dimension-5 operator 
\begin{equation}
    {\cal L}_{\rm axion} = g_{a\gamma\gamma}{\bf E}\cdot{\bf B}\, a,
\end{equation}
where ${\bf E} \equiv \mathbf{E}_{\rm eff}$ is the electric field in the crystal lattice, ${\bf B} \equiv \mathbf{B}_{\rm FEL}$ is the magnetic field associated with the electromagnetic wave of the X-ray photon, $a$ is the $CP$-conserving  scalar field of the axion, and $g_{a\gamma\gamma}$ is the axion-photon coupling. Note that here and throughout this manuscript unless otherwise noted, natural Heaviside-Lorentz units are used. 

Experiments employing the above coupling exploit the Primakoff effect viz. that there is a finite probability for a photon to decay into an axion in the presence of another photon, typically given by a static, external field. The conversion (or regeneration) probability is maximized when the electric and magnetic fields of these two photons are aligned. This probability increases linearly with interaction length. 

\begin{figure}
    \centering
    \includegraphics[width=\linewidth]{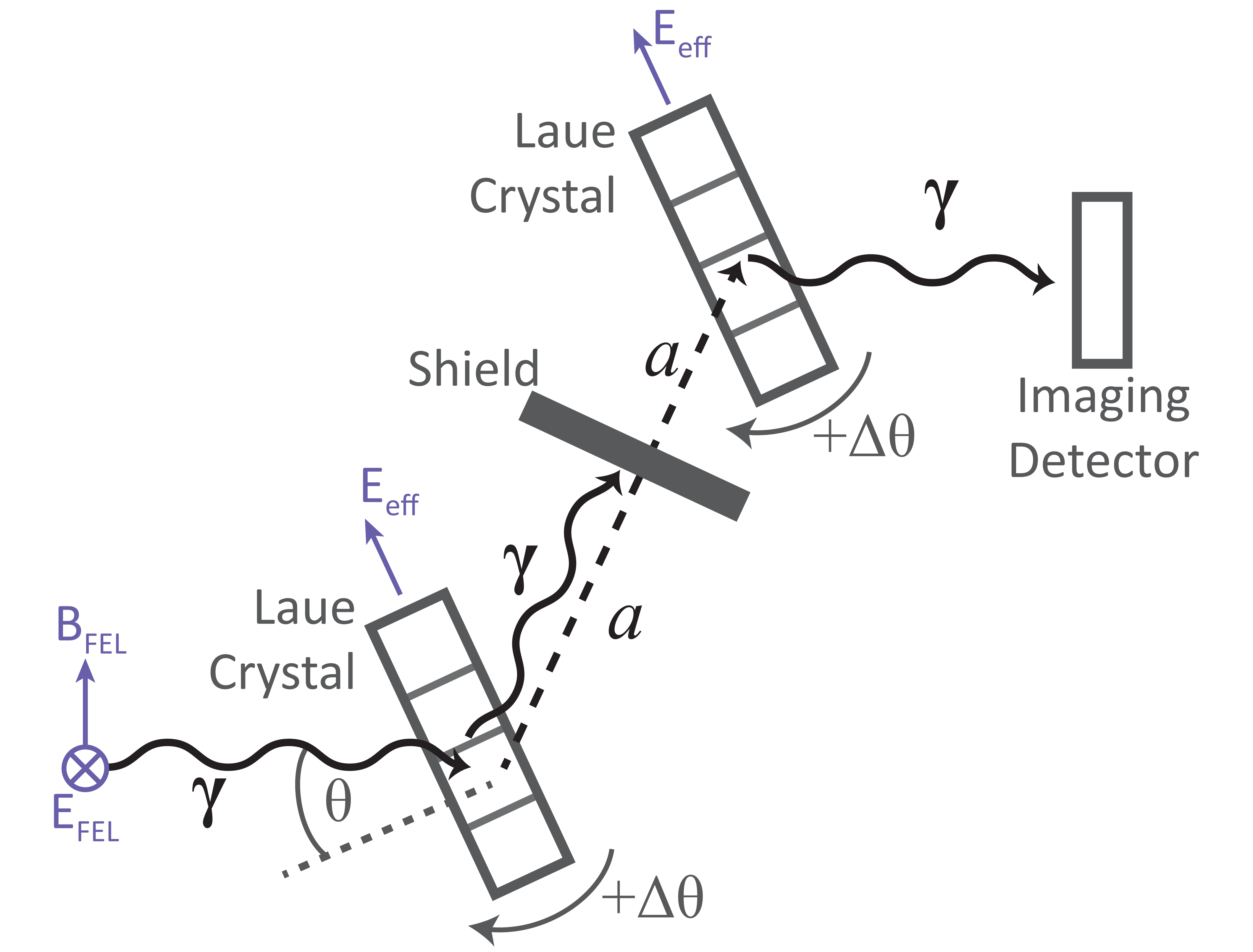}
    \caption{Diagram of the setup in our experiment; the X-ray beam propagates from left to right. Here $\mathbf{B}_{\rm FEL}$ is  the magnetic field in the XFEL beam and $\mathbf{E}_{\rm eff}$ is the crystaline electric field. Axion production and photon regeneration are expected to take place via the effective electric field within a pair of monolithic crystals, Ge (220) in Laue geometry, with dimensions: $10 \;\si{\milli\meter} \times 10  \;\si{\milli\meter}\times 0.5 \;\si{\milli\meter}$. A pair of piezoelectric rotation stages (Xeryon, XRT-U30) were used to orient the germanium crystals. The shield is a \SI{1}{\milli\meter} thick titanium sheet. The polarisation of the X-ray beam maximises the value of $\mathbf{B}_{\rm FEL}\cdot\mathbf{E}_{\rm eff}$ and thus the probability of axion production.}
    \label{fig:setup}
\end{figure}


Our experimental setup is depicted in Figure~\ref{fig:setup}. It shows two germanium (Ge) crystals oriented in Laue geometry, with their lattice planes parallel to one-another. The $\sigma$-polarised XFEL beam impinges on the first crystal from the left. The angle between the wave-vector of the incoming X-ray beam and the lattice planes in the crystals is denoted $\theta$. An important detail is that the Laue geometry is preferable to the more conventional Bragg scattering geometry because of the Borrmann effect, through which the transmission of X-rays in the Laue case is increased \cite{Kovev1969,Buchmuller:1989rb,Liao:2010ig,Yamaji:2017pep}. 


Both axions and Laue diffracted photons are transmitted through the first crystal. These are denoted respectively by $a$ and $\gamma$ in the figure. The photons are absorbed by a radiation shield but the weakly interacting axions impinge on the second crystal. Here the strong electric field enables the regeneration of photons via the inverse Primakoff process. These regenerated photons are observed by a detector downstream of the crystals. In the configuration where $\theta=\theta_\mathrm{B}$ (here $\theta_\mathrm{B}$ is the Bragg angle), the design is sensitive to a broad range of axion mass $m_a$  satisfying the inequality 
\begin{equation}
    |m_a^2 - m_\gamma^2| \lesssim \frac{4k_\gamma}{L_\mathrm{eff}},
\end{equation}
where $m_\gamma=\SI{44}{\electronvolt}$ is the plasma frequency of the valence electrons in the conversion crystals \cite{Yamaji:2017pep}; $k_\gamma$ is the photon energy; and $L_\mathrm{eff}$ is the effective path length of X-rays within a crystal. We use units where $\hbar=c=1$. 

In the case where there is a detuning from the Bragg angle, by $\Delta\theta=\theta-\theta_\mathrm{B}$, it can be shown \cite{Liao:2010ig,Yamaji:2017pep} that this setup becomes sensitive to a narrow range of axion mass ($\Delta m_a \sim 10^{-3}\;\si{\electronvolt}$) centered on 
\begin{equation}
    m_a = \sqrt{m_\gamma^2 + 2q_T k_\gamma \mathrm{cos}(\theta_\mathrm{B}) \Delta\theta},
\end{equation}
where $q_T=\SI{6.20}{\kilo\electronvolt}$ is the magnitude of the reciprocal lattice vector. This means that by sweeping through different values of $\Delta\theta$ it is possible to search for heavy axions with mass in the interval between the plasma frequency of the crystal and the projection of the incoming photon energy onto the reciprocal lattice vector. 

The EuXFEL was operated in a seeded mode, with \SI{9.8}{\kilo\electronvolt} photon energy (wavelength, $\lambda_{\rm x}=2 \pi/ k_\gamma = 1.265$ \AA). The repetition rate was \SI{10}{\hertz}, with one pulse per train. The X-ray beam was collimated by upstream compound refractive lenses (CRLs). The full-width-half-maximum of the beam transverse profile was measured to be $\SI{400}{\micro\meter}$ at the center of the interaction chamber.  
The axion-photon conversion probability $P(a\leftrightarrow \gamma)$ 
for Laue-case diffraction is given by \cite{Yamaji:2017pep}
\begin{equation}
  P(a\leftrightarrow \gamma) =  \left(\frac{1}{4}g_{a\gamma\gamma} E_{\rm eff} \,L_\mathrm{eff}\,\cos\theta_\mathrm{B}  \right)^2,
  \label{gagg0}
\end{equation}
where $E_{\rm eff} = 7.3\times 10^{10} $ \si{\volt\per\meter} is the crystalline electric field \cite{Yamaji:2017pep}, 
and 
\begin{equation}
    L_\mathrm{eff} = 2 L_{\rm att}^B \left(1-{\rm e}^{-L_{\rm x}/2 L_{\rm att}^B}\right),
\end{equation}
where $L_{\rm x}=\ell /\cos(\theta_B+\Delta\theta)$ is the X-ray path length inside the crystals ($\ell=500$ \SI{}{\micro\meter} is the thickness of each crystal) and $L_{\rm att}^B = 1499.8$ \SI{}{\micro\meter} (for $\sigma$-polarization) \footnote{From https://x-server.gmca.aps.anl.gov/x0h.html.}.

Since the X-ray pulse duration in our experiment is short compared to that at a synchrotron facility, the result presented above requires a modification. For a short (i.e. transform-limited) X-ray pulse, the width of the rocking curve ($\Delta \theta_{RC}$) and timescale of the scattering process ($\Delta t$) form a time-bandwidth product given by \cite{wark,Shvydko:2012rzc}:
\begin{equation}
    \Delta\theta_{RC} \Delta t \simeq \frac{\lambda_{\rm x}\tan\theta_B}{c},
\end{equation}
where for clarity we have reinstated $c$, the speed of light.

Because of the Borrmann effect, the extinction length of the X-rays is longer than the X-ray path-length in the crystal and therefore the characteristic timescale is simply given by the geometric time-delay due to scattering off multiple planes,
\begin{equation}
    c \Delta t = 2\ell\tan\theta_B\sin\theta_B.
\end{equation}
Combining these two expressions yields a rocking curve width $\Delta \theta_{RC} \simeq 0.4$ \textmu rad, which is far narrower than the Darwin width, $\Delta \theta_D = 44$ \textmu rad for Ge $(220)$ \footnote{From https://x-server.gmca.aps.anl.gov/x0h.html.}. 
By the rocking curve width, we mean the \emph{actual} angular spread in incoming X-rays which are transmitted through the crystal. Meanwhile, the Darwin width refers to the \emph{predicted} angular spread obtained with a theory that neglects time dependence. 

As shown in Ref.~\cite{Yamaji:2017pep}, the effective conversion length is inversely proportional to the width of the rocking curve. However, in deriving Eq.~(\ref{gagg0}), it was
assumed that the Darwin width and rocking curve width were equivalent. This narrowing of the rocking curve may also be interpreted as a change in the effective index of refraction inside the crystal lattice. Following the same steps as in Ref.~\cite{Yamaji:2017pep}, it can then be readily shown that the interaction amplitude must increase by a factor $\xi_B = \Delta\theta_D / \Delta \theta_{RC}$. Thus, the scattering probability becomes
\begin{equation}
  P(a\leftrightarrow \gamma) \simeq  \left(\frac{1}{4}g_{a\gamma\gamma} E_{\rm eff} \,L_\mathrm{eff}\,\xi_B\,\cos\theta_\mathrm{B}  \right)^2.
  \label{gagg}
\end{equation}
While an exact derivation of this result would require a full solution of the time-dependent dynamical diffraction equations, as outlined in Ref.~\cite{wark}, the above expression is accurate to within a factor of order unity.

{\it Results.}
The regenerated photons were measured using a silicon hybrid-pixel JUNGFRAU detector \cite{Mozzanica2016}. Further details regarding data acquisition and the steps taken to ensure alignment stability are provided in the supplementary material for this manuscript.

Our search was limited to 5 discrete $\Delta \theta$ values, with data collected for $60-90 \, \si{\min}$ at each angle. Table \ref{tab:res}, shows the bounds on the axion-photon coupling determined from our data at each detuning angle. 
\begin{table}
    \centering
    \begin{tabular}{|c|c|c|c|c|}
    \hline
    $\Delta \theta \; [\si{\milli\radian}]$ & $m_a \; [\si{\electronvolt}]$ & $N_{\rm in} \;(\times 10^{16})$ & $g_{a \gamma \gamma} \; [\si{\per \giga \electronvolt}$]\\
    \hline
    $0.0$              & $\lesssim  44$    & $2.6$ & $3.91\times 10^{-4}$ \\ 
    $1.0$& $3.4\times 10^2$  & $2.4$ & $3.10\times 10^{-4}$ \\
    $1.8$& $4.6\times 10^2$  & $1.6$ & $3.87\times 10^{-4}$ \\
    $10.0$& $1.1\times 10^3$  & $1.7$ & $3.69\times 10^{-4}$ \\
    $50.0$& $2.4\times 10^3$  & $1.5$ & $2.76\times 10^{-4}$ \\
    \hline 
    \end{tabular}
    \caption{Summary of the different runs which were performed during the experiment. The detuning angles, $\Delta\theta$; corresponding masses, $m_a$; total number of photons incident upon the apparatus, $N_{\rm in}$; and inferred bound on the strength of the axion-photon coupling constant,  $g_{a \gamma \gamma}$ are indicated.}
    \label{tab:res}
\end{table}

\begin{figure}
    \centering
    \includegraphics[width=0.9\linewidth]{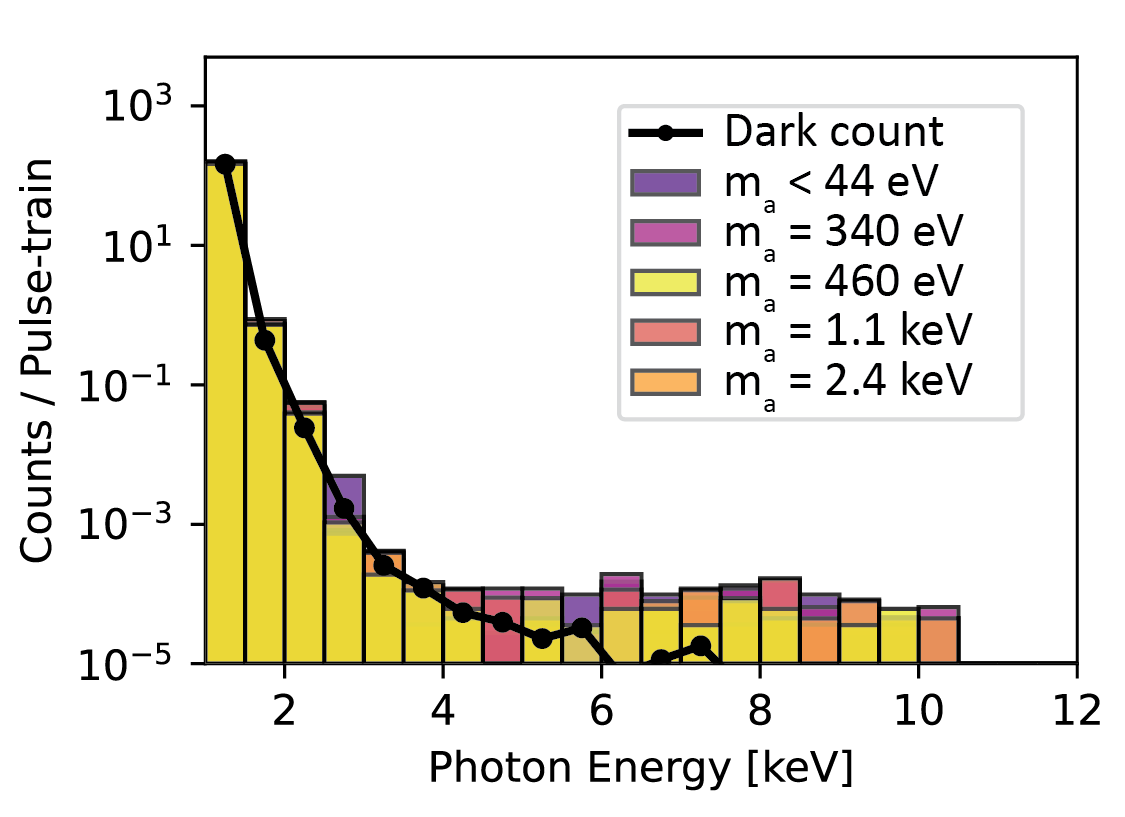}
    \caption{A histogram showing the events detected across all acquisitions and over the whole detector area. The number of counts in a 24 hour dark run are also shown.}
    \label{fig:hist}
\end{figure}

Figure~\ref{fig:hist} is a histogram which shows energy-resolved events for each of the data sets which are detailed in Table \ref{tab:res}. These are compared against the number of counts in a 24-hour long dark-run.
ALP observations can be distinguished from the background as reconverted X-rays must be identical to the primary EuXFEL X-rays, and moreover must fall inside the region on the detector which is impacted by the X-ray beam when the shield is absent.

\begin{figure}
    \centering
    \includegraphics[width=1.1\linewidth]{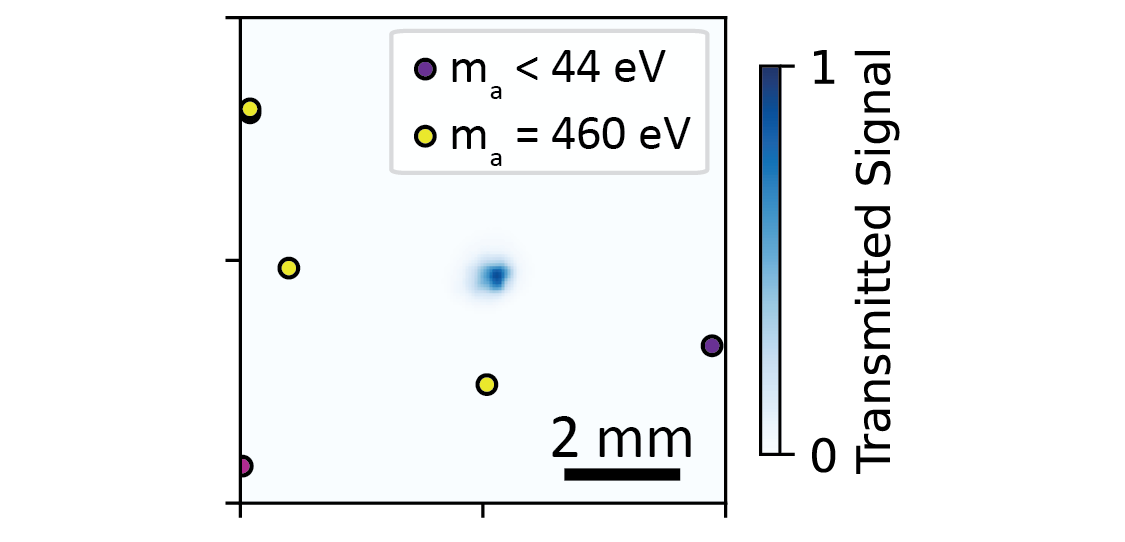}
    \caption{An image showing the transmitted signal in the absence of the radiation shield (blue colour-map) overlaid with the position of $k_\gamma \geq \SI{4}{\kilo\electronvolt}$ events across all data acquisitions. A fiducial indicating scale on the detector plane is also shown.}
    \label{fig:hits}
\end{figure}

To establish if any of the  few events in the relevant energy band do fall upon the X-ray spot and might therefore be associated with axion production, hit-maps of events were produced as in Figure~\ref{fig:hits}. The blue colour map shows transmission through the setup in the absence of the radiation shield while the overlaid data points indicate the location of hits on the detector with a photon energy exceeding \SI{4}{\kilo\electronvolt} for each of the data sets in Table \ref{tab:res}. 

As Figure~\ref{fig:hits} shows, there are no events which overlap with the region of the X-ray spot (the darker blue region in the center of the figure). Their absence implies that no events consistent with axion production were detected during the experiment.     
The corresponding limit on the axion-photon coupling is then obtained by inverting Eq.~(\ref{gagg}):
\begin{equation}
    g_{a\gamma\gamma} < 
    \left(\frac{1}{4}\,{E_{\rm eff}\, L_{\rm B}\,\xi_B\,\cos \theta_\mathrm{B}}\right)^{-1} P(a\leftrightarrow \gamma)^{1/2},
    \label{eq:bound}
\end{equation}
with $P(a\leftrightarrow \gamma)^2=\left({N_{\rm det}} / {\eta N_{\rm in}} \right)$; $N_{\rm det}$ is the detected number of photons; $N_{\rm in}$ is the number of input photons. 

The efficiency factor $\eta$ accounts for losses associated with the deviation from parallelism between the two crystals; fluctuations in the exact X-ray energy; and the quantum efficiency of the detector. 
The value of $\eta$ was obtained experimentally: At the beginning and end of each data run, the crystals were tuned to the Bragg angle and the radiation shield was removed in order to characterise the experimental setup. During these characterisation phases, the efficiency factor for the $i$-th run at a given detuning angle, $\eta_i$, was given by   
\begin{equation}
    \eta_i = \frac{1}{T_{\rm Ge}^2}  \frac{E^{\rm JF, ch}_{i}} {E^{\rm in, ch}_{i}},
\end{equation}
where $T_{\rm Ge}$ is the transmission factor associated with a single crystal; $E^{\rm JF, ch}_i$ is the total X-ray dose measured on the (downstream) JUNGFRAU detector during these characterisation phases; and $E^{\rm in, aq}_i$ is the total X-ray dose measured (during characterisation) on a passive upstream monitor \cite{Maltezopoulos2019}.

\begin{figure}
    \centering
    \includegraphics[width=0.9\linewidth]{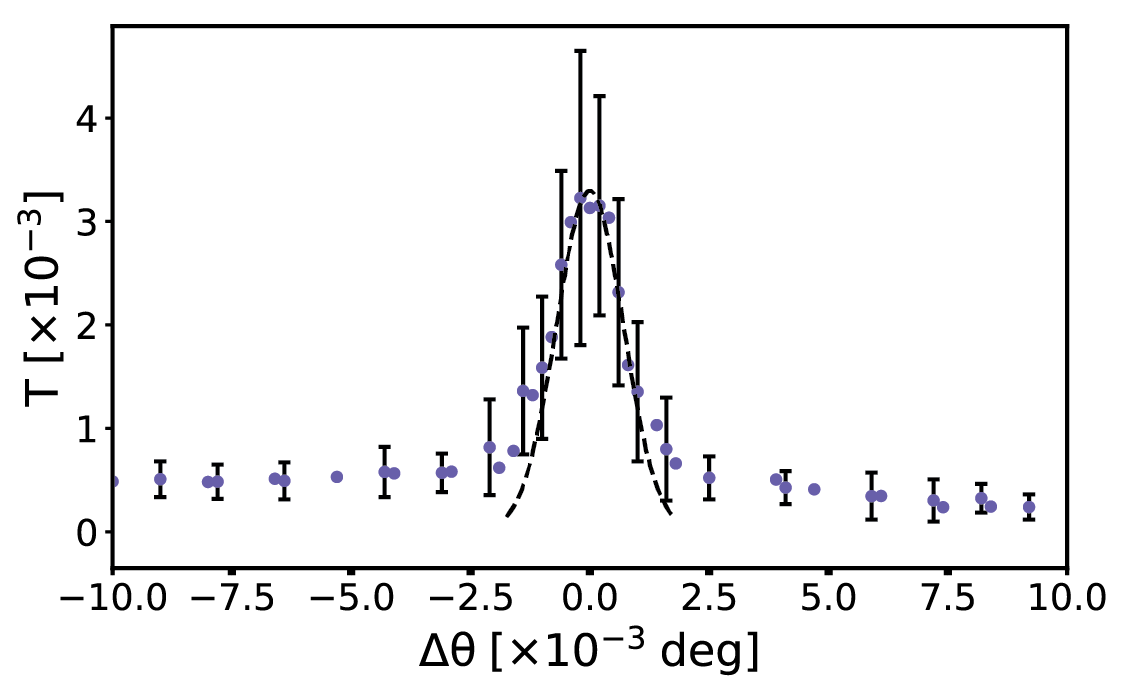}
    \caption{X-ray transmission through a single Ge crystal as a function of the detuning angle $\Delta \theta$. The central peak is fitted with a Gaussian (dashed line) of width $\Delta \theta_s \approx 17.4$ \textmu rad. An average of 145 shots per angular point are used to construct the peak curve, while 32 shots are used for each angular point on the baseline. The error bars on the measurements are 1$\sigma$.}
    \label{fig:rock}
\end{figure}

Because of the very narrow rocking curve for Laue-case diffraction, a single Ge crystal can be used to determine the EuXFEL spectral profile by detuning it from the Bragg angle and recording the transmitted intensity on a separate JUNGFRAU detector as a function of the detuning angle. This is shown in Figure~\ref{fig:rock}, where the seeded X-ray beam is shown to have an energy bandwidth of
$\Delta E/E = \Delta \theta_s/\tan \theta_B = 5.2 \times 10^{-5}$, or $\sim$0.5 eV at 9.8 keV. This is indeed expected for a self-seeded beam \cite{Emma:2017qcq}, and the variations in the trasmitted intensity are associated with shot-to-shot variability in the exact seeded pulse energy. Overall, the transmission through a single crystal is determined to be of order $T_{\rm Ge} \approx 3\times 10^{-3}$. 

For the data collection phases of a given dataset, the value of $\eta N_{\rm in}$ was then taken to be  
\begin{equation}
    \eta N_{\rm in} = \sum_i \eta_i E^{\rm in, aq}_i / k_\gamma,
\end{equation}
where the summation is across all runs at a given detuning angle; $E^{\rm in, aq}_i$ is the dose measured on the passive upstream monitor during data collection; and $k_\gamma=\SI{9.8}{\kilo\electronvolt}$ is the photon energy. 
To obtain a 90\% CL upper bound based on the observation of zero events consistent with axion production, we then take $N_{\rm det}=2.3$ events~\cite{Junk:1999kv}.

\begin{figure}
    \centering
    \includegraphics[width=\linewidth]{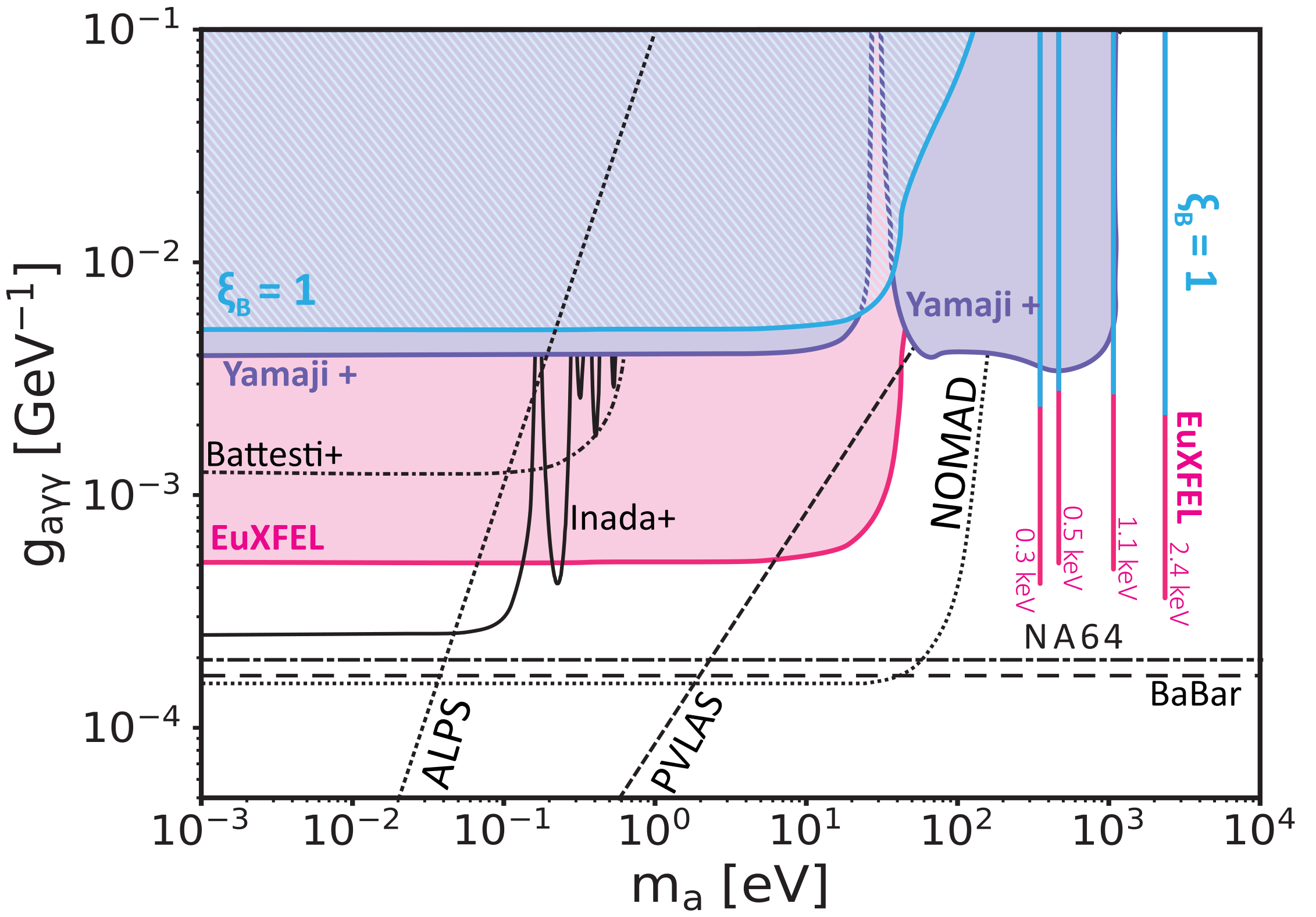}
    \caption{Bounds on the axion-photon coupling from our experiment (pink), compared with those from Yamaji et al. \cite{Yamaji:2017pep,Yamaji:2018ufo} (purple). The excluded region  (blue) taking $\xi_B=1$ in Eq.~(\ref{eq:bound}) is also shown to illustrate the improvement due to the higher photon number. Shown for comparison are bounds from other laboratory searches: NOMAD \cite{NOMAD:2000usb}, PVLAS \cite{DellaValle:2014wea}, ALPS \cite{Ehret:2010mh}, NA64 \cite{NA64:2020qwq}, BaBar \cite{BaBar:2017tiz,Dolan:2017osp}, Battesti et al. \cite{Battesti:2010dm} and Inada et al. \cite{Inada:2016jzh}. To aid visualisation, the width of the off-Bragg bounds inferred from our search (in reality only $\sim10^{-3}\; \si{\electronvolt}$) is exaggerated here. The masses associated with off-Bragg searches are also labelled in the figure.}
    \label{fig:bounds}
\end{figure}

{\it Concluding remarks}.

The outcome of this analysis of data collected at EuXFEL is shown in Figure~\ref{fig:bounds}, which summarizes bounds in the meV--few keV mass range, from searches for laboratory-generated axions. We were able to improve on the results from Ref.~\cite{Yamaji:2018ufo} at several discrete axion masses. 

For $m_a \gtrsim \SI{200}{\electronvolt}$, we are able achieve a sensitivity within a factor 10 of the most competitive previous searches, namely NA64 \cite{NA64:2020qwq} and BaBar \cite{BaBar:2017tiz,Dolan:2017osp}. Although our search is presently not as sensitive as these experiments, our result constitute an important validation, especially as the NA64 and BaBar limits were extracted assuming different production/detection mechanisms for axions, namely spontaneous axion decay and/or flavor-changing meson decay, rather than the Primakoff process as in our case.

We emphasise that this is not the best sensitivity achievable with the present setup. Issues with X-ray heating forced us to attenuate the X-ray flux by a factor of $10^3$. Moreover, the X-ray bunch structure was set with the number of pulses per train limited to $1$, out of a possible $300$. Issues with retaining alignment also limited data acquisition time to $60-90\;\si{\min}$ at each detuning angle; with a more stable setup that would include active cooling of the first conversion crystal, these times could be increased by a factor of 30. 
Furthermore, we could also fully exploit the Borrmann effect and use Ge crystals up to 1.5 mm in thickness. 
Taken together these improvements would 
increase the sensitivity by a factor $\sim 150$,
bringing the estimated bounds down to $2\times 10^{-6}$ GeV$^{-1}$, close to the expectation for QCD axions to be dark matter \cite{OHare:2024nmr}.
Below $\sim$1 eV, these bounds are also comparable to proposed photon regeneration experiments using superconducting pulsed magnetic fields \cite{Rabadan:2005dm}. 
Currently no experiment has the level of sensitivity for $m_a \gtrsim 10$~eV that we anticipate for future searches with the platform described here.

\section*{Acknowledgements}
This research received funding from the UK Engineering and Physical Sciences Research Council (EP/X01133X/1 \& EP/X010791/1). SS and GG belong to the ‘Quantum Sensors for the Hidden Sector’ consortium funded by the UK Science \& Technology Facilities Council (ST/T006277/1) and wish thank Andreas Ringwald for helpful comments. JH was partially supported by the US Defence Threat Reduction Agency (HDTRA1-20-1-0001) and the US Department of Energy (DE-SC0020434 \& DE-NA0003764). CH was partly funded through the UK XFEL Physical Sciences Hub.
We acknowledge the European XFEL in Schenefeld, Germany for the provision of beamtime at HED SASE2 and thank all the staff for their assistance. 

\section*{Data Availability}
The raw data recorded for this experiment at the European XFEL are available at \href{https://doi.org/10.22003/XFEL.EU-DATA-003326-00}{DOI:10.22003/XFEL.EU-DATA-003326-00}.
\bibliography{ref}

\begin{thebibliography}{59}%
\makeatletter
\providecommand \@ifxundefined [1]{%
 \@ifx{#1\undefined}
}%
\providecommand \@ifnum [1]{%
 \ifnum #1\expandafter \@firstoftwo
 \else \expandafter \@secondoftwo
 \fi
}%
\providecommand \@ifx [1]{%
 \ifx #1\expandafter \@firstoftwo
 \else \expandafter \@secondoftwo
 \fi
}%
\providecommand \natexlab [1]{#1}%
\providecommand \enquote  [1]{``#1''}%
\providecommand \bibnamefont  [1]{#1}%
\providecommand \bibfnamefont [1]{#1}%
\providecommand \citenamefont [1]{#1}%
\providecommand \href@noop [0]{\@secondoftwo}%
\providecommand \href [0]{\begingroup \@sanitize@url \@href}%
\providecommand \@href[1]{\@@startlink{#1}\@@href}%
\providecommand \@@href[1]{\endgroup#1\@@endlink}%
\providecommand \@sanitize@url [0]{\catcode `\\12\catcode `\$12\catcode
  `\&12\catcode `\#12\catcode `\^12\catcode `\_12\catcode `\%12\relax}%
\providecommand \@@startlink[1]{}%
\providecommand \@@endlink[0]{}%
\providecommand \url  [0]{\begingroup\@sanitize@url \@url }%
\providecommand \@url [1]{\endgroup\@href {#1}{\urlprefix }}%
\providecommand \urlprefix  [0]{URL }%
\providecommand \Eprint [0]{\href }%
\providecommand \doibase [0]{https://doi.org/}%
\providecommand \selectlanguage [0]{\@gobble}%
\providecommand \bibinfo  [0]{\@secondoftwo}%
\providecommand \bibfield  [0]{\@secondoftwo}%
\providecommand \translation [1]{[#1]}%
\providecommand \BibitemOpen [0]{}%
\providecommand \bibitemStop [0]{}%
\providecommand \bibitemNoStop [0]{.\EOS\space}%
\providecommand \EOS [0]{\spacefactor3000\relax}%
\providecommand \BibitemShut  [1]{\csname bibitem#1\endcsname}%
\let\auto@bib@innerbib\@empty
\bibitem [{\citenamefont {Yamaji}\ \emph {et~al.}(2018)\citenamefont {Yamaji},
  \citenamefont {Tamasaku}, \citenamefont {Namba}, \citenamefont {Yamazaki},\
  and\ \citenamefont {Seino}}]{Yamaji:2018ufo}%
  \BibitemOpen
  \bibfield  {author} {\bibinfo {author} {\bibfnamefont {T.}~\bibnamefont
  {Yamaji}}, \bibinfo {author} {\bibfnamefont {K.}~\bibnamefont {Tamasaku}},
  \bibinfo {author} {\bibfnamefont {T.}~\bibnamefont {Namba}}, \bibinfo
  {author} {\bibfnamefont {T.}~\bibnamefont {Yamazaki}},\ and\ \bibinfo
  {author} {\bibfnamefont {Y.}~\bibnamefont {Seino}},\ }\bibfield  {title}
  {\bibinfo {title} {{Search for Axion like particles using Laue-case
  conversion in a single crystal}},\ }\href
  {https://doi.org/10.1016/j.physletb.2018.05.068} {\bibfield  {journal}
  {\bibinfo  {journal} {Phys. Lett. B}\ }\textbf {\bibinfo {volume} {782}},\
  \bibinfo {pages} {523} (\bibinfo {year} {2018})},\ \Eprint
  {https://arxiv.org/abs/1802.08388} {arXiv:1802.08388 [hep-ex]} \BibitemShut
  {NoStop}%
\bibitem [{\citenamefont {Peccei}\ and\ \citenamefont
  {Quinn}(1977)}]{Peccei:1977hh}%
  \BibitemOpen
  \bibfield  {author} {\bibinfo {author} {\bibfnamefont {R.~D.}\ \bibnamefont
  {Peccei}}\ and\ \bibinfo {author} {\bibfnamefont {H.~R.}\ \bibnamefont
  {Quinn}},\ }\bibfield  {title} {\bibinfo {title} {{CP Conservation in the
  Presence of Instantons}},\ }\href
  {https://doi.org/10.1103/PhysRevLett.38.1440} {\bibfield  {journal} {\bibinfo
   {journal} {Phys. Rev. Lett.}\ }\textbf {\bibinfo {volume} {38}},\ \bibinfo
  {pages} {1440} (\bibinfo {year} {1977})}\BibitemShut {NoStop}%
\bibitem [{\citenamefont {Weinberg}(1978)}]{Weinberg:1977ma}%
  \BibitemOpen
  \bibfield  {author} {\bibinfo {author} {\bibfnamefont {S.}~\bibnamefont
  {Weinberg}},\ }\bibfield  {title} {\bibinfo {title} {{A New Light Boson?}},\
  }\href {https://doi.org/10.1103/PhysRevLett.40.223} {\bibfield  {journal}
  {\bibinfo  {journal} {Phys. Rev. Lett.}\ }\textbf {\bibinfo {volume} {40}},\
  \bibinfo {pages} {223} (\bibinfo {year} {1978})}\BibitemShut {NoStop}%
\bibitem [{\citenamefont {Wilczek}(1978)}]{Wilczek:1977pj}%
  \BibitemOpen
  \bibfield  {author} {\bibinfo {author} {\bibfnamefont {F.}~\bibnamefont
  {Wilczek}},\ }\bibfield  {title} {\bibinfo {title} {{Problem of Strong $P$
  and $T$ Invariance in the Presence of Instantons}},\ }\href
  {https://doi.org/10.1103/PhysRevLett.40.279} {\bibfield  {journal} {\bibinfo
  {journal} {Phys. Rev. Lett.}\ }\textbf {\bibinfo {volume} {40}},\ \bibinfo
  {pages} {279} (\bibinfo {year} {1978})}\BibitemShut {NoStop}%
\bibitem [{\citenamefont {Svrcek}\ and\ \citenamefont
  {Witten}(2006)}]{Svrcek:2006yi}%
  \BibitemOpen
  \bibfield  {author} {\bibinfo {author} {\bibfnamefont {P.}~\bibnamefont
  {Svrcek}}\ and\ \bibinfo {author} {\bibfnamefont {E.}~\bibnamefont
  {Witten}},\ }\bibfield  {title} {\bibinfo {title} {{Axions In String
  Theory}},\ }\href {https://doi.org/10.1088/1126-6708/2006/06/051} {\bibfield
  {journal} {\bibinfo  {journal} {JHEP}\ }\textbf {\bibinfo {volume} {06}},\
  \bibinfo {pages} {051}},\ \Eprint {https://arxiv.org/abs/hep-th/0605206}
  {arXiv:hep-th/0605206} \BibitemShut {NoStop}%
\bibitem [{\citenamefont {Preskill}\ \emph {et~al.}(1983)\citenamefont
  {Preskill}, \citenamefont {Wise},\ and\ \citenamefont
  {Wilczek}}]{Preskill:1982cy}%
  \BibitemOpen
  \bibfield  {author} {\bibinfo {author} {\bibfnamefont {J.}~\bibnamefont
  {Preskill}}, \bibinfo {author} {\bibfnamefont {M.~B.}\ \bibnamefont {Wise}},\
  and\ \bibinfo {author} {\bibfnamefont {F.}~\bibnamefont {Wilczek}},\
  }\bibfield  {title} {\bibinfo {title} {{Cosmology of the Invisible Axion}},\
  }\href {https://doi.org/10.1016/0370-2693(83)90637-8} {\bibfield  {journal}
  {\bibinfo  {journal} {Phys. Lett. B}\ }\textbf {\bibinfo {volume} {120}},\
  \bibinfo {pages} {127} (\bibinfo {year} {1983})}\BibitemShut {NoStop}%
\bibitem [{\citenamefont {Abbott}\ and\ \citenamefont
  {Sikivie}(1983)}]{Abbott:1982af}%
  \BibitemOpen
  \bibfield  {author} {\bibinfo {author} {\bibfnamefont {L.~F.}\ \bibnamefont
  {Abbott}}\ and\ \bibinfo {author} {\bibfnamefont {P.}~\bibnamefont
  {Sikivie}},\ }\bibfield  {title} {\bibinfo {title} {{A Cosmological Bound on
  the Invisible Axion}},\ }\href {https://doi.org/10.1016/0370-2693(83)90638-X}
  {\bibfield  {journal} {\bibinfo  {journal} {Phys. Lett. B}\ }\textbf
  {\bibinfo {volume} {120}},\ \bibinfo {pages} {133} (\bibinfo {year}
  {1983})}\BibitemShut {NoStop}%
\bibitem [{\citenamefont {Dine}\ and\ \citenamefont
  {Fischler}(1983)}]{Dine:1982ah}%
  \BibitemOpen
  \bibfield  {author} {\bibinfo {author} {\bibfnamefont {M.}~\bibnamefont
  {Dine}}\ and\ \bibinfo {author} {\bibfnamefont {W.}~\bibnamefont
  {Fischler}},\ }\bibfield  {title} {\bibinfo {title} {{The Not So Harmless
  Axion}},\ }\href {https://doi.org/10.1016/0370-2693(83)90639-1} {\bibfield
  {journal} {\bibinfo  {journal} {Phys. Lett. B}\ }\textbf {\bibinfo {volume}
  {120}},\ \bibinfo {pages} {137} (\bibinfo {year} {1983})}\BibitemShut
  {NoStop}%
\bibitem [{\citenamefont {Sikivie}(1983)}]{Sikivie:1983ip}%
  \BibitemOpen
  \bibfield  {author} {\bibinfo {author} {\bibfnamefont {P.}~\bibnamefont
  {Sikivie}},\ }\bibfield  {title} {\bibinfo {title} {{Experimental Tests of
  the Invisible Axion}},\ }\href {https://doi.org/10.1103/PhysRevLett.51.1415}
  {\bibfield  {journal} {\bibinfo  {journal} {Phys. Rev. Lett.}\ }\textbf
  {\bibinfo {volume} {51}},\ \bibinfo {pages} {1415} (\bibinfo {year}
  {1983})},\ \bibinfo {note} {[Erratum: Phys.Rev.Lett. 52, 695
  (1984)]}\BibitemShut {NoStop}%
\bibitem [{\citenamefont {Semertzidis}\ and\ \citenamefont
  {Youn}(2022)}]{Semertzidis:2021rxs}%
  \BibitemOpen
  \bibfield  {author} {\bibinfo {author} {\bibfnamefont {Y.~K.}\ \bibnamefont
  {Semertzidis}}\ and\ \bibinfo {author} {\bibfnamefont {S.}~\bibnamefont
  {Youn}},\ }\bibfield  {title} {\bibinfo {title} {{Axion dark matter: How to
  see it?}},\ }\href {https://doi.org/10.1126/sciadv.abm9928} {\bibfield
  {journal} {\bibinfo  {journal} {Sci. Adv.}\ }\textbf {\bibinfo {volume}
  {8}},\ \bibinfo {pages} {abm9928} (\bibinfo {year} {2022})},\ \Eprint
  {https://arxiv.org/abs/2104.14831} {arXiv:2104.14831 [hep-ph]} \BibitemShut
  {NoStop}%
\bibitem [{\citenamefont {Kim}(1979)}]{Kim:1979if}%
  \BibitemOpen
  \bibfield  {author} {\bibinfo {author} {\bibfnamefont {J.~E.}\ \bibnamefont
  {Kim}},\ }\bibfield  {title} {\bibinfo {title} {{Weak Interaction Singlet and
  Strong CP Invariance}},\ }\href {https://doi.org/10.1103/PhysRevLett.43.103}
  {\bibfield  {journal} {\bibinfo  {journal} {Phys. Rev. Lett.}\ }\textbf
  {\bibinfo {volume} {43}},\ \bibinfo {pages} {103} (\bibinfo {year}
  {1979})}\BibitemShut {NoStop}%
\bibitem [{\citenamefont {Dine}\ \emph {et~al.}(1981)\citenamefont {Dine},
  \citenamefont {Fischler},\ and\ \citenamefont {Srednicki}}]{Dine:1981rt}%
  \BibitemOpen
  \bibfield  {author} {\bibinfo {author} {\bibfnamefont {M.}~\bibnamefont
  {Dine}}, \bibinfo {author} {\bibfnamefont {W.}~\bibnamefont {Fischler}},\
  and\ \bibinfo {author} {\bibfnamefont {M.}~\bibnamefont {Srednicki}},\
  }\bibfield  {title} {\bibinfo {title} {{A Simple Solution to the Strong CP
  Problem with a Harmless Axion}},\ }\href
  {https://doi.org/10.1016/0370-2693(81)90590-6} {\bibfield  {journal}
  {\bibinfo  {journal} {Phys. Lett. B}\ }\textbf {\bibinfo {volume} {104}},\
  \bibinfo {pages} {199} (\bibinfo {year} {1981})}\BibitemShut {NoStop}%
\bibitem [{\citenamefont {Zhitnitsky}(1980)}]{Zhitnitsky:1980tq}%
  \BibitemOpen
  \bibfield  {author} {\bibinfo {author} {\bibfnamefont {A.~R.}\ \bibnamefont
  {Zhitnitsky}},\ }\bibfield  {title} {\bibinfo {title} {{On Possible
  Suppression of the Axion Hadron Interactions. (In Russian)}},\ }\href@noop {}
  {\bibfield  {journal} {\bibinfo  {journal} {Sov. J. Nucl. Phys.}\ }\textbf
  {\bibinfo {volume} {31}},\ \bibinfo {pages} {260} (\bibinfo {year}
  {1980})}\BibitemShut {NoStop}%
\bibitem [{\citenamefont {Ringwald}\ and\ \citenamefont
  {Saikawa}(2016)}]{Ringwald:2015dsf}%
  \BibitemOpen
  \bibfield  {author} {\bibinfo {author} {\bibfnamefont {A.}~\bibnamefont
  {Ringwald}}\ and\ \bibinfo {author} {\bibfnamefont {K.}~\bibnamefont
  {Saikawa}},\ }\bibfield  {title} {\bibinfo {title} {{Axion dark matter in the
  post-inflationary Peccei-Quinn symmetry breaking scenario}},\ }\href
  {https://doi.org/10.1103/PhysRevD.93.085031} {\bibfield  {journal} {\bibinfo
  {journal} {Phys. Rev. D}\ }\textbf {\bibinfo {volume} {93}},\ \bibinfo
  {pages} {085031} (\bibinfo {year} {2016})},\ \bibinfo {note} {[Addendum:
  Phys.Rev.D 94, 049908 (2016)]},\ \Eprint {https://arxiv.org/abs/1512.06436}
  {arXiv:1512.06436 [hep-ph]} \BibitemShut {NoStop}%
\bibitem [{\citenamefont {Beyer}\ and\ \citenamefont
  {Sarkar}(2023)}]{Beyer:2022ywc}%
  \BibitemOpen
  \bibfield  {author} {\bibinfo {author} {\bibfnamefont {K.~A.}\ \bibnamefont
  {Beyer}}\ and\ \bibinfo {author} {\bibfnamefont {S.}~\bibnamefont {Sarkar}},\
  }\bibfield  {title} {\bibinfo {title} {{Ruling out light axions: The writing
  is on the wall}},\ }\href {https://doi.org/10.21468/SciPostPhys.15.1.003}
  {\bibfield  {journal} {\bibinfo  {journal} {SciPost Phys.}\ }\textbf
  {\bibinfo {volume} {15}},\ \bibinfo {pages} {003} (\bibinfo {year} {2023})},\
  \Eprint {https://arxiv.org/abs/2211.14635} {arXiv:2211.14635 [hep-ph]}
  \BibitemShut {NoStop}%
\bibitem [{\citenamefont {Kamionkowski}\ and\ \citenamefont
  {March-Russell}(1992)}]{Kamionkowski:1992mf}%
  \BibitemOpen
  \bibfield  {author} {\bibinfo {author} {\bibfnamefont {M.}~\bibnamefont
  {Kamionkowski}}\ and\ \bibinfo {author} {\bibfnamefont {J.}~\bibnamefont
  {March-Russell}},\ }\bibfield  {title} {\bibinfo {title} {{Planck scale
  physics and the Peccei-Quinn mechanism}},\ }\href
  {https://doi.org/10.1016/0370-2693(92)90492-M} {\bibfield  {journal}
  {\bibinfo  {journal} {Phys. Lett. B}\ }\textbf {\bibinfo {volume} {282}},\
  \bibinfo {pages} {137} (\bibinfo {year} {1992})},\ \Eprint
  {https://arxiv.org/abs/hep-th/9202003} {arXiv:hep-th/9202003} \BibitemShut
  {NoStop}%
\bibitem [{\citenamefont {Holman}\ \emph {et~al.}(1992)\citenamefont {Holman},
  \citenamefont {Hsu}, \citenamefont {Kephart}, \citenamefont {Kolb},
  \citenamefont {Watkins},\ and\ \citenamefont {Widrow}}]{Holman:1992us}%
  \BibitemOpen
  \bibfield  {author} {\bibinfo {author} {\bibfnamefont {R.}~\bibnamefont
  {Holman}}, \bibinfo {author} {\bibfnamefont {S.~D.~H.}\ \bibnamefont {Hsu}},
  \bibinfo {author} {\bibfnamefont {T.~W.}\ \bibnamefont {Kephart}}, \bibinfo
  {author} {\bibfnamefont {E.~W.}\ \bibnamefont {Kolb}}, \bibinfo {author}
  {\bibfnamefont {R.}~\bibnamefont {Watkins}},\ and\ \bibinfo {author}
  {\bibfnamefont {L.~M.}\ \bibnamefont {Widrow}},\ }\bibfield  {title}
  {\bibinfo {title} {{Solutions to the strong CP problem in a world with
  gravity}},\ }\href {https://doi.org/10.1016/0370-2693(92)90491-L} {\bibfield
  {journal} {\bibinfo  {journal} {Phys. Lett. B}\ }\textbf {\bibinfo {volume}
  {282}},\ \bibinfo {pages} {132} (\bibinfo {year} {1992})},\ \Eprint
  {https://arxiv.org/abs/hep-ph/9203206} {arXiv:hep-ph/9203206} \BibitemShut
  {NoStop}%
\bibitem [{\citenamefont {Barr}\ and\ \citenamefont
  {Seckel}(1992)}]{Barr:1992qq}%
  \BibitemOpen
  \bibfield  {author} {\bibinfo {author} {\bibfnamefont {S.~M.}\ \bibnamefont
  {Barr}}\ and\ \bibinfo {author} {\bibfnamefont {D.}~\bibnamefont {Seckel}},\
  }\bibfield  {title} {\bibinfo {title} {{Planck scale corrections to axion
  models}},\ }\href {https://doi.org/10.1103/PhysRevD.46.539} {\bibfield
  {journal} {\bibinfo  {journal} {Phys. Rev. D}\ }\textbf {\bibinfo {volume}
  {46}},\ \bibinfo {pages} {539} (\bibinfo {year} {1992})}\BibitemShut
  {NoStop}%
\bibitem [{\citenamefont {Caputo}\ and\ \citenamefont
  {Raffelt}(2024)}]{Caputo:2024oqc}%
  \BibitemOpen
  \bibfield  {author} {\bibinfo {author} {\bibfnamefont {A.}~\bibnamefont
  {Caputo}}\ and\ \bibinfo {author} {\bibfnamefont {G.}~\bibnamefont
  {Raffelt}},\ }\bibfield  {title} {\bibinfo {title} {{Astrophysical Axion
  Bounds: The 2024 Edition}},\ }\href {https://doi.org/10.22323/1.454.0041}
  {\bibfield  {journal} {\bibinfo  {journal} {PoS}\ }\textbf {\bibinfo {volume}
  {COSMICWISPers}},\ \bibinfo {pages} {041} (\bibinfo {year} {2024})},\ \Eprint
  {https://arxiv.org/abs/2401.13728} {arXiv:2401.13728 [hep-ph]} \BibitemShut
  {NoStop}%
\bibitem [{\citenamefont {Anastassopoulos}\ \emph {et~al.}(2017)\citenamefont
  {Anastassopoulos} \emph {et~al.}}]{CAST:2017uph}%
  \BibitemOpen
  \bibfield  {author} {\bibinfo {author} {\bibfnamefont {V.}~\bibnamefont
  {Anastassopoulos}} \emph {et~al.} (\bibinfo {collaboration} {CAST}),\
  }\bibfield  {title} {\bibinfo {title} {{New CAST Limit on the Axion-Photon
  Interaction}},\ }\href {https://doi.org/10.1038/nphys4109} {\bibfield
  {journal} {\bibinfo  {journal} {Nature Phys.}\ }\textbf {\bibinfo {volume}
  {13}},\ \bibinfo {pages} {584} (\bibinfo {year} {2017})},\ \Eprint
  {https://arxiv.org/abs/1705.02290} {arXiv:1705.02290 [hep-ex]} \BibitemShut
  {NoStop}%
\bibitem [{\citenamefont {Avignone}\ \emph {et~al.}(1998)\citenamefont
  {Avignone} \emph {et~al.}}]{SOLAX:1997lpz}%
  \BibitemOpen
  \bibfield  {author} {\bibinfo {author} {\bibfnamefont {F.~T.}\ \bibnamefont
  {Avignone}, \bibfnamefont {III}} \emph {et~al.} (\bibinfo {collaboration}
  {SOLAX}),\ }\bibfield  {title} {\bibinfo {title} {{Experimental search for
  solar axions via coherent Primakoff conversion in a germanium
  spectrometer}},\ }\href {https://doi.org/10.1103/PhysRevLett.81.5068}
  {\bibfield  {journal} {\bibinfo  {journal} {Phys. Rev. Lett.}\ }\textbf
  {\bibinfo {volume} {81}},\ \bibinfo {pages} {5068} (\bibinfo {year}
  {1998})},\ \Eprint {https://arxiv.org/abs/astro-ph/9708008}
  {arXiv:astro-ph/9708008} \BibitemShut {NoStop}%
\bibitem [{\citenamefont {Bernabei}\ \emph {et~al.}(2001)\citenamefont
  {Bernabei} \emph {et~al.}}]{Bernabei:2001ny}%
  \BibitemOpen
  \bibfield  {author} {\bibinfo {author} {\bibfnamefont {R.}~\bibnamefont
  {Bernabei}} \emph {et~al.},\ }\bibfield  {title} {\bibinfo {title} {{Search
  for solar axions by Primakoff effect in NaI crystals}},\ }\href
  {https://doi.org/10.1016/S0370-2693(01)00840-1} {\bibfield  {journal}
  {\bibinfo  {journal} {Phys. Lett. B}\ }\textbf {\bibinfo {volume} {515}},\
  \bibinfo {pages} {6} (\bibinfo {year} {2001})}\BibitemShut {NoStop}%
\bibitem [{\citenamefont {Morales}\ \emph {et~al.}(2002)\citenamefont {Morales}
  \emph {et~al.}}]{COSME:2001jci}%
  \BibitemOpen
  \bibfield  {author} {\bibinfo {author} {\bibfnamefont {A.}~\bibnamefont
  {Morales}} \emph {et~al.} (\bibinfo {collaboration} {COSME}),\ }\bibfield
  {title} {\bibinfo {title} {{Particle dark matter and solar axion searches
  with a small germanium detector at the Canfranc Underground Laboratory}},\
  }\href {https://doi.org/10.1016/S0927-6505(01)00117-7} {\bibfield  {journal}
  {\bibinfo  {journal} {Astropart. Phys.}\ }\textbf {\bibinfo {volume} {16}},\
  \bibinfo {pages} {325} (\bibinfo {year} {2002})},\ \Eprint
  {https://arxiv.org/abs/hep-ex/0101037} {arXiv:hep-ex/0101037} \BibitemShut
  {NoStop}%
\bibitem [{\citenamefont {Ahmed}\ \emph {et~al.}(2009)\citenamefont {Ahmed}
  \emph {et~al.}}]{CDMS:2009fba}%
  \BibitemOpen
  \bibfield  {author} {\bibinfo {author} {\bibfnamefont {Z.}~\bibnamefont
  {Ahmed}} \emph {et~al.} (\bibinfo {collaboration} {CDMS}),\ }\bibfield
  {title} {\bibinfo {title} {{Search for Axions with the CDMS Experiment}},\
  }\href {https://doi.org/10.1103/PhysRevLett.103.141802} {\bibfield  {journal}
  {\bibinfo  {journal} {Phys. Rev. Lett.}\ }\textbf {\bibinfo {volume} {103}},\
  \bibinfo {pages} {141802} (\bibinfo {year} {2009})},\ \Eprint
  {https://arxiv.org/abs/0902.4693} {arXiv:0902.4693 [hep-ex]} \BibitemShut
  {NoStop}%
\bibitem [{\citenamefont {Belli}\ \emph {et~al.}(2012)\citenamefont {Belli},
  \citenamefont {Bernabei}, \citenamefont {Cappella}, \citenamefont {Cerulli},
  \citenamefont {Danevich}, \citenamefont {Incicchitti}, \citenamefont
  {Kobychev}, \citenamefont {Laubenstein}, \citenamefont {Polischuk},\ and\
  \citenamefont {Tretyak}}]{Belli:2012zz}%
  \BibitemOpen
  \bibfield  {author} {\bibinfo {author} {\bibfnamefont {P.}~\bibnamefont
  {Belli}}, \bibinfo {author} {\bibfnamefont {R.}~\bibnamefont {Bernabei}},
  \bibinfo {author} {\bibfnamefont {F.}~\bibnamefont {Cappella}}, \bibinfo
  {author} {\bibfnamefont {R.}~\bibnamefont {Cerulli}}, \bibinfo {author}
  {\bibfnamefont {F.~A.}\ \bibnamefont {Danevich}}, \bibinfo {author}
  {\bibfnamefont {A.}~\bibnamefont {Incicchitti}}, \bibinfo {author}
  {\bibfnamefont {V.~V.}\ \bibnamefont {Kobychev}}, \bibinfo {author}
  {\bibfnamefont {M.}~\bibnamefont {Laubenstein}}, \bibinfo {author}
  {\bibfnamefont {O.~G.}\ \bibnamefont {Polischuk}},\ and\ \bibinfo {author}
  {\bibfnamefont {V.~I.}\ \bibnamefont {Tretyak}},\ }\bibfield  {title}
  {\bibinfo {title} {{Search for Li-7 solar axions using resonant absorption in
  LiF crystal: Final results}},\ }\href
  {https://doi.org/10.1016/j.physletb.2012.03.067} {\bibfield  {journal}
  {\bibinfo  {journal} {Phys. Lett. B}\ }\textbf {\bibinfo {volume} {711}},\
  \bibinfo {pages} {41} (\bibinfo {year} {2012})}\BibitemShut {NoStop}%
\bibitem [{\citenamefont {Armengaud}\ \emph {et~al.}(2013)\citenamefont
  {Armengaud} \emph {et~al.}}]{Armengaud:2013rta}%
  \BibitemOpen
  \bibfield  {author} {\bibinfo {author} {\bibfnamefont {E.}~\bibnamefont
  {Armengaud}} \emph {et~al.},\ }\bibfield  {title} {\bibinfo {title} {{Axion
  searches with the EDELWEISS-II experiment}},\ }\href
  {https://doi.org/10.1088/1475-7516/2013/11/067} {\bibfield  {journal}
  {\bibinfo  {journal} {JCAP}\ }\textbf {\bibinfo {volume} {11}},\ \bibinfo
  {pages} {067}},\ \Eprint {https://arxiv.org/abs/1307.1488} {arXiv:1307.1488
  [astro-ph.CO]} \BibitemShut {NoStop}%
\bibitem [{\citenamefont {Arnquist}\ \emph {et~al.}(2022)\citenamefont
  {Arnquist} \emph {et~al.}}]{Majorana:2022bse}%
  \BibitemOpen
  \bibfield  {author} {\bibinfo {author} {\bibfnamefont {I.~J.}\ \bibnamefont
  {Arnquist}} \emph {et~al.} (\bibinfo {collaboration} {Majorana}),\ }\bibfield
   {title} {\bibinfo {title} {{Search for Solar Axions via Axion-Photon
  Coupling with the Majorana Demonstrator}},\ }\href
  {https://doi.org/10.1103/PhysRevLett.129.081803} {\bibfield  {journal}
  {\bibinfo  {journal} {Phys. Rev. Lett.}\ }\textbf {\bibinfo {volume} {129}},\
  \bibinfo {pages} {081803} (\bibinfo {year} {2022})},\ \Eprint
  {https://arxiv.org/abs/2206.05789} {arXiv:2206.05789 [nucl-ex]} \BibitemShut
  {NoStop}%
\bibitem [{\citenamefont {Aprile}\ \emph {et~al.}(2022)\citenamefont {Aprile}
  \emph {et~al.}}]{XENON:2022ltv}%
  \BibitemOpen
  \bibfield  {author} {\bibinfo {author} {\bibfnamefont {E.}~\bibnamefont
  {Aprile}} \emph {et~al.} (\bibinfo {collaboration} {XENON}),\ }\bibfield
  {title} {\bibinfo {title} {{Search for New Physics in Electronic Recoil Data
  from XENONnT}},\ }\href {https://doi.org/10.1103/PhysRevLett.129.161805}
  {\bibfield  {journal} {\bibinfo  {journal} {Phys. Rev. Lett.}\ }\textbf
  {\bibinfo {volume} {129}},\ \bibinfo {pages} {161805} (\bibinfo {year}
  {2022})},\ \Eprint {https://arxiv.org/abs/2207.11330} {arXiv:2207.11330
  [hep-ex]} \BibitemShut {NoStop}%
\bibitem [{\citenamefont {Dent}\ \emph {et~al.}(2023)\citenamefont {Dent},
  \citenamefont {Dutta},\ and\ \citenamefont {Thompson}}]{Dent:2023gzl}%
  \BibitemOpen
  \bibfield  {author} {\bibinfo {author} {\bibfnamefont {J.~B.}\ \bibnamefont
  {Dent}}, \bibinfo {author} {\bibfnamefont {B.}~\bibnamefont {Dutta}},\ and\
  \bibinfo {author} {\bibfnamefont {A.}~\bibnamefont {Thompson}},\ }\bibfield
  {title} {\bibinfo {title} {{Bragg-Primakoff Axion Photoconversion in Crystal
  Detectors}},\ }\href@noop {} {\  (\bibinfo {year} {2023})},\ \Eprint
  {https://arxiv.org/abs/2307.04861} {arXiv:2307.04861 [hep-ph]} \BibitemShut
  {NoStop}%
\bibitem [{\citenamefont {Jaeckel}\ \emph {et~al.}(2007)\citenamefont
  {Jaeckel}, \citenamefont {Masso}, \citenamefont {Redondo}, \citenamefont
  {Ringwald},\ and\ \citenamefont {Takahashi}}]{Jaeckel:2006xm}%
  \BibitemOpen
  \bibfield  {author} {\bibinfo {author} {\bibfnamefont {J.}~\bibnamefont
  {Jaeckel}}, \bibinfo {author} {\bibfnamefont {E.}~\bibnamefont {Masso}},
  \bibinfo {author} {\bibfnamefont {J.}~\bibnamefont {Redondo}}, \bibinfo
  {author} {\bibfnamefont {A.}~\bibnamefont {Ringwald}},\ and\ \bibinfo
  {author} {\bibfnamefont {F.}~\bibnamefont {Takahashi}},\ }\bibfield  {title}
  {\bibinfo {title} {{The Need for purely laboratory-based axion-like particle
  searches}},\ }\href {https://doi.org/10.1103/PhysRevD.75.013004} {\bibfield
  {journal} {\bibinfo  {journal} {Phys. Rev. D}\ }\textbf {\bibinfo {volume}
  {75}},\ \bibinfo {pages} {013004} (\bibinfo {year} {2007})},\ \Eprint
  {https://arxiv.org/abs/hep-ph/0610203} {arXiv:hep-ph/0610203} \BibitemShut
  {NoStop}%
\bibitem [{\citenamefont {Bar}\ \emph {et~al.}(2020)\citenamefont {Bar},
  \citenamefont {Blum},\ and\ \citenamefont {D'Amico}}]{Bar:2019ifz}%
  \BibitemOpen
  \bibfield  {author} {\bibinfo {author} {\bibfnamefont {N.}~\bibnamefont
  {Bar}}, \bibinfo {author} {\bibfnamefont {K.}~\bibnamefont {Blum}},\ and\
  \bibinfo {author} {\bibfnamefont {G.}~\bibnamefont {D'Amico}},\ }\bibfield
  {title} {\bibinfo {title} {{Is there a supernova bound on axions?}},\ }\href
  {https://doi.org/10.1103/PhysRevD.101.123025} {\bibfield  {journal} {\bibinfo
   {journal} {Phys. Rev. D}\ }\textbf {\bibinfo {volume} {101}},\ \bibinfo
  {pages} {123025} (\bibinfo {year} {2020})},\ \Eprint
  {https://arxiv.org/abs/1907.05020} {arXiv:1907.05020 [hep-ph]} \BibitemShut
  {NoStop}%
\bibitem [{\citenamefont {Astier}\ \emph {et~al.}(2000)\citenamefont {Astier}
  \emph {et~al.}}]{NOMAD:2000usb}%
  \BibitemOpen
  \bibfield  {author} {\bibinfo {author} {\bibfnamefont {P.}~\bibnamefont
  {Astier}} \emph {et~al.} (\bibinfo {collaboration} {NOMAD}),\ }\bibfield
  {title} {\bibinfo {title} {{Search for eV (pseudo)scalar penetrating
  particles in the SPS neutrino beam}},\ }\href
  {https://doi.org/10.1016/S0370-2693(00)00375-0} {\bibfield  {journal}
  {\bibinfo  {journal} {Phys. Lett. B}\ }\textbf {\bibinfo {volume} {479}},\
  \bibinfo {pages} {371} (\bibinfo {year} {2000})}\BibitemShut {NoStop}%
\bibitem [{\citenamefont {Lees}\ \emph {et~al.}(2017)\citenamefont {Lees} \emph
  {et~al.}}]{BaBar:2017tiz}%
  \BibitemOpen
  \bibfield  {author} {\bibinfo {author} {\bibfnamefont {J.~P.}\ \bibnamefont
  {Lees}} \emph {et~al.} (\bibinfo {collaboration} {BaBar}),\ }\bibfield
  {title} {\bibinfo {title} {{Search for Invisible Decays of a Dark Photon
  Produced in ${e}^{+}{e}^{-}$ Collisions at BaBar}},\ }\href
  {https://doi.org/10.1103/PhysRevLett.119.131804} {\bibfield  {journal}
  {\bibinfo  {journal} {Phys. Rev. Lett.}\ }\textbf {\bibinfo {volume} {119}},\
  \bibinfo {pages} {131804} (\bibinfo {year} {2017})},\ \Eprint
  {https://arxiv.org/abs/1702.03327} {arXiv:1702.03327 [hep-ex]} \BibitemShut
  {NoStop}%
\bibitem [{\citenamefont {Dolan}\ \emph {et~al.}(2017)\citenamefont {Dolan},
  \citenamefont {Ferber}, \citenamefont {Hearty}, \citenamefont {Kahlhoefer},\
  and\ \citenamefont {Schmidt-Hoberg}}]{Dolan:2017osp}%
  \BibitemOpen
  \bibfield  {author} {\bibinfo {author} {\bibfnamefont {M.~J.}\ \bibnamefont
  {Dolan}}, \bibinfo {author} {\bibfnamefont {T.}~\bibnamefont {Ferber}},
  \bibinfo {author} {\bibfnamefont {C.}~\bibnamefont {Hearty}}, \bibinfo
  {author} {\bibfnamefont {F.}~\bibnamefont {Kahlhoefer}},\ and\ \bibinfo
  {author} {\bibfnamefont {K.}~\bibnamefont {Schmidt-Hoberg}},\ }\bibfield
  {title} {\bibinfo {title} {{Revised constraints and Belle II sensitivity for
  visible and invisible axion-like particles}},\ }\href
  {https://doi.org/10.1007/JHEP12(2017)094} {\bibfield  {journal} {\bibinfo
  {journal} {JHEP}\ }\textbf {\bibinfo {volume} {12}},\ \bibinfo {pages}
  {094}},\ \bibinfo {note} {[Erratum: JHEP 03, 190 (2021)]},\ \Eprint
  {https://arxiv.org/abs/1709.00009} {arXiv:1709.00009 [hep-ph]} \BibitemShut
  {NoStop}%
\bibitem [{\citenamefont {Banerjee}\ \emph {et~al.}(2020)\citenamefont
  {Banerjee} \emph {et~al.}}]{NA64:2020qwq}%
  \BibitemOpen
  \bibfield  {author} {\bibinfo {author} {\bibfnamefont {D.}~\bibnamefont
  {Banerjee}} \emph {et~al.} (\bibinfo {collaboration} {NA64}),\ }\bibfield
  {title} {\bibinfo {title} {{Search for Axionlike and Scalar Particles with
  the NA64 Experiment}},\ }\href
  {https://doi.org/10.1103/PhysRevLett.125.081801} {\bibfield  {journal}
  {\bibinfo  {journal} {Phys. Rev. Lett.}\ }\textbf {\bibinfo {volume} {125}},\
  \bibinfo {pages} {081801} (\bibinfo {year} {2020})},\ \Eprint
  {https://arxiv.org/abs/2005.02710} {arXiv:2005.02710 [hep-ex]} \BibitemShut
  {NoStop}%
\bibitem [{\citenamefont {Zastrau}\ \emph {et~al.}(2021)\citenamefont
  {Zastrau}, \citenamefont {Appel}, \citenamefont {Baehtz}, \citenamefont
  {Baehr}, \citenamefont {Batchelor}, \citenamefont {Bergh{\"{a}}user},
  \citenamefont {Banjafar}, \citenamefont {Brambrink}, \citenamefont
  {Cerantola}, \citenamefont {Cowan} \emph {et~al.}}]{Zastrau2021}%
  \BibitemOpen
  \bibfield  {author} {\bibinfo {author} {\bibfnamefont {U.}~\bibnamefont
  {Zastrau}}, \bibinfo {author} {\bibfnamefont {K.}~\bibnamefont {Appel}},
  \bibinfo {author} {\bibfnamefont {C.}~\bibnamefont {Baehtz}}, \bibinfo
  {author} {\bibfnamefont {O.}~\bibnamefont {Baehr}}, \bibinfo {author}
  {\bibfnamefont {L.}~\bibnamefont {Batchelor}}, \bibinfo {author}
  {\bibfnamefont {A.}~\bibnamefont {Bergh{\"{a}}user}}, \bibinfo {author}
  {\bibfnamefont {M.}~\bibnamefont {Banjafar}}, \bibinfo {author}
  {\bibfnamefont {E.}~\bibnamefont {Brambrink}}, \bibinfo {author}
  {\bibfnamefont {V.}~\bibnamefont {Cerantola}}, \bibinfo {author}
  {\bibfnamefont {T.~E.}\ \bibnamefont {Cowan}}, \emph {et~al.},\ }\bibfield
  {title} {\bibinfo {title} {{The High Energy Density Scientific Instrument at
  the European XFEL}},\ }\href {https://doi.org/10.1107/S1600577521007335}
  {\bibfield  {journal} {\bibinfo  {journal} {Journal of Synchrotron
  Radiation}\ }\textbf {\bibinfo {volume} {28}},\ \bibinfo {pages} {1393}
  (\bibinfo {year} {2021})}\BibitemShut {NoStop}%
\bibitem [{\citenamefont {Robilliard}\ \emph {et~al.}(2007)\citenamefont
  {Robilliard}, \citenamefont {Battesti}, \citenamefont {Fouche}, \citenamefont
  {Mauchain}, \citenamefont {Sautivet}, \citenamefont {Amiranoff},\ and\
  \citenamefont {Rizzo}}]{Robilliard:2007bq}%
  \BibitemOpen
  \bibfield  {author} {\bibinfo {author} {\bibfnamefont {C.}~\bibnamefont
  {Robilliard}}, \bibinfo {author} {\bibfnamefont {R.}~\bibnamefont
  {Battesti}}, \bibinfo {author} {\bibfnamefont {M.}~\bibnamefont {Fouche}},
  \bibinfo {author} {\bibfnamefont {J.}~\bibnamefont {Mauchain}}, \bibinfo
  {author} {\bibfnamefont {A.-M.}\ \bibnamefont {Sautivet}}, \bibinfo {author}
  {\bibfnamefont {F.}~\bibnamefont {Amiranoff}},\ and\ \bibinfo {author}
  {\bibfnamefont {C.}~\bibnamefont {Rizzo}},\ }\bibfield  {title} {\bibinfo
  {title} {{No light shining through a wall}},\ }\href
  {https://doi.org/10.1103/PhysRevLett.99.190403} {\bibfield  {journal}
  {\bibinfo  {journal} {Phys. Rev. Lett.}\ }\textbf {\bibinfo {volume} {99}},\
  \bibinfo {pages} {190403} (\bibinfo {year} {2007})},\ \Eprint
  {https://arxiv.org/abs/0707.1296} {arXiv:0707.1296 [hep-ex]} \BibitemShut
  {NoStop}%
\bibitem [{\citenamefont {Ehret}\ \emph {et~al.}(2010)\citenamefont {Ehret}
  \emph {et~al.}}]{Ehret:2010mh}%
  \BibitemOpen
  \bibfield  {author} {\bibinfo {author} {\bibfnamefont {K.}~\bibnamefont
  {Ehret}} \emph {et~al.},\ }\bibfield  {title} {\bibinfo {title} {{New ALPS
  Results on Hidden-Sector Lightweights}},\ }\href
  {https://doi.org/10.1016/j.physletb.2010.04.066} {\bibfield  {journal}
  {\bibinfo  {journal} {Phys. Lett. B}\ }\textbf {\bibinfo {volume} {689}},\
  \bibinfo {pages} {149} (\bibinfo {year} {2010})},\ \Eprint
  {https://arxiv.org/abs/1004.1313} {arXiv:1004.1313 [hep-ex]} \BibitemShut
  {NoStop}%
\bibitem [{\citenamefont {Ballou}\ \emph {et~al.}(2015)\citenamefont {Ballou}
  \emph {et~al.}}]{OSQAR:2015qdv}%
  \BibitemOpen
  \bibfield  {author} {\bibinfo {author} {\bibfnamefont {R.}~\bibnamefont
  {Ballou}} \emph {et~al.} (\bibinfo {collaboration} {OSQAR}),\ }\bibfield
  {title} {\bibinfo {title} {{New exclusion limits on scalar and pseudoscalar
  axionlike particles from light shining through a wall}},\ }\href
  {https://doi.org/10.1103/PhysRevD.92.092002} {\bibfield  {journal} {\bibinfo
  {journal} {Phys. Rev. D}\ }\textbf {\bibinfo {volume} {92}},\ \bibinfo
  {pages} {092002} (\bibinfo {year} {2015})},\ \Eprint
  {https://arxiv.org/abs/1506.08082} {arXiv:1506.08082 [hep-ex]} \BibitemShut
  {NoStop}%
\bibitem [{\citenamefont {Doyle}\ and\ \citenamefont
  {Turner}(1968)}]{Doyle1968}%
  \BibitemOpen
  \bibfield  {author} {\bibinfo {author} {\bibfnamefont {P.~A.}\ \bibnamefont
  {Doyle}}\ and\ \bibinfo {author} {\bibfnamefont {P.~S.}\ \bibnamefont
  {Turner}},\ }\href {https://doi.org/10.1107/S0567739468000756} {\bibinfo
  {title} {{Relativistic Hartree{--}Fock X-ray and electron scattering
  factors}}} (\bibinfo {year} {1968})\BibitemShut {NoStop}%
\bibitem [{\citenamefont {Baier}\ \emph {et~al.}(1998)\citenamefont {Baier},
  \citenamefont {Katkov},\ and\ \citenamefont {Strakhovenko}}]{Baier1998}%
  \BibitemOpen
  \bibfield  {author} {\bibinfo {author} {\bibfnamefont {V.~N.}\ \bibnamefont
  {Baier}}, \bibinfo {author} {\bibfnamefont {V.~M.}\ \bibnamefont {Katkov}},\
  and\ \bibinfo {author} {\bibfnamefont {V.}~\bibnamefont {Strakhovenko}},\
  }\href {https://doi.org/10.1142/2216} {\emph {\bibinfo {title}
  {{Electromagnetic Processes at High Energies in Oriented Single Crystals}}}}\
  (\bibinfo  {publisher} {World Scientific},\ \bibinfo {year}
  {1998})\BibitemShut {NoStop}%
\bibitem [{\citenamefont {Atkinson}\ \emph {et~al.}(1982)\citenamefont
  {Atkinson}, \citenamefont {Bak}, \citenamefont {Bussey}, \citenamefont
  {Christensen}, \citenamefont {Ellison}, \citenamefont {Ellison},
  \citenamefont {Eriksen}, \citenamefont {Giddings}, \citenamefont
  {Hughes-Jones}, \citenamefont {Marsh}, \citenamefont {Mercer}, \citenamefont
  {Meyer}, \citenamefont {Møller}, \citenamefont {Newton}, \citenamefont
  {Pavlopoulos}, \citenamefont {Sharp}, \citenamefont {Stensgaard},
  \citenamefont {Suffert},\ and\ \citenamefont {Uggerhøj}}]{Atkinson1982}%
  \BibitemOpen
  \bibfield  {author} {\bibinfo {author} {\bibfnamefont {M.}~\bibnamefont
  {Atkinson}}, \bibinfo {author} {\bibfnamefont {J.}~\bibnamefont {Bak}},
  \bibinfo {author} {\bibfnamefont {P.}~\bibnamefont {Bussey}}, \bibinfo
  {author} {\bibfnamefont {P.}~\bibnamefont {Christensen}}, \bibinfo {author}
  {\bibfnamefont {J.}~\bibnamefont {Ellison}}, \bibinfo {author} {\bibfnamefont
  {R.}~\bibnamefont {Ellison}}, \bibinfo {author} {\bibfnamefont
  {K.}~\bibnamefont {Eriksen}}, \bibinfo {author} {\bibfnamefont
  {D.}~\bibnamefont {Giddings}}, \bibinfo {author} {\bibfnamefont
  {R.}~\bibnamefont {Hughes-Jones}}, \bibinfo {author} {\bibfnamefont
  {B.}~\bibnamefont {Marsh}}, \bibinfo {author} {\bibfnamefont
  {D.}~\bibnamefont {Mercer}}, \bibinfo {author} {\bibfnamefont
  {F.}~\bibnamefont {Meyer}}, \bibinfo {author} {\bibfnamefont
  {S.}~\bibnamefont {Møller}}, \bibinfo {author} {\bibfnamefont
  {D.}~\bibnamefont {Newton}}, \bibinfo {author} {\bibfnamefont
  {P.}~\bibnamefont {Pavlopoulos}}, \bibinfo {author} {\bibfnamefont
  {P.}~\bibnamefont {Sharp}}, \bibinfo {author} {\bibfnamefont
  {R.}~\bibnamefont {Stensgaard}}, \bibinfo {author} {\bibfnamefont
  {M.}~\bibnamefont {Suffert}},\ and\ \bibinfo {author} {\bibfnamefont
  {E.}~\bibnamefont {Uggerhøj}},\ }\bibfield  {title} {\bibinfo {title}
  {Radiation from planar channeled 5–55 solgevc positrons and elctrons},\
  }\href {https://doi.org/https://doi.org/10.1016/0370-2693(82)91027-9}
  {\bibfield  {journal} {\bibinfo  {journal} {Physics Letters B}\ }\textbf
  {\bibinfo {volume} {110}},\ \bibinfo {pages} {162} (\bibinfo {year}
  {1982})}\BibitemShut {NoStop}%
\bibitem [{\citenamefont {Buchmuller}\ and\ \citenamefont
  {Hoogeveen}(1990)}]{Buchmuller:1989rb}%
  \BibitemOpen
  \bibfield  {author} {\bibinfo {author} {\bibfnamefont {W.}~\bibnamefont
  {Buchmuller}}\ and\ \bibinfo {author} {\bibfnamefont {F.}~\bibnamefont
  {Hoogeveen}},\ }\bibfield  {title} {\bibinfo {title} {{Coherent Production of
  Light Scalar Particles in Bragg Scattering}},\ }\href
  {https://doi.org/10.1016/0370-2693(90)91444-G} {\bibfield  {journal}
  {\bibinfo  {journal} {Phys. Lett. B}\ }\textbf {\bibinfo {volume} {237}},\
  \bibinfo {pages} {278} (\bibinfo {year} {1990})}\BibitemShut {NoStop}%
\bibitem [{\citenamefont {Yamaji}\ \emph {et~al.}(2017)\citenamefont {Yamaji},
  \citenamefont {Yamazaki}, \citenamefont {Tamasaku},\ and\ \citenamefont
  {Namba}}]{Yamaji:2017pep}%
  \BibitemOpen
  \bibfield  {author} {\bibinfo {author} {\bibfnamefont {T.}~\bibnamefont
  {Yamaji}}, \bibinfo {author} {\bibfnamefont {T.}~\bibnamefont {Yamazaki}},
  \bibinfo {author} {\bibfnamefont {K.}~\bibnamefont {Tamasaku}},\ and\
  \bibinfo {author} {\bibfnamefont {T.}~\bibnamefont {Namba}},\ }\bibfield
  {title} {\bibinfo {title} {{Theoretical calculation of coherent Laue-case
  conversion between x-rays and ALPs for an x-ray light-shining-through-a-wall
  experiment}},\ }\href {https://doi.org/10.1103/PhysRevD.96.115001} {\bibfield
   {journal} {\bibinfo  {journal} {Phys. Rev. D}\ }\textbf {\bibinfo {volume}
  {96}},\ \bibinfo {pages} {115001} (\bibinfo {year} {2017})},\ \Eprint
  {https://arxiv.org/abs/1709.03299} {arXiv:1709.03299 [physics.ins-det]}
  \BibitemShut {NoStop}%
\bibitem [{\citenamefont {Kovev}\ \emph {et~al.}(1969)\citenamefont {Kovev},
  \citenamefont {Efimov},\ and\ \citenamefont {Korovin}}]{Kovev1969}%
  \BibitemOpen
  \bibfield  {author} {\bibinfo {author} {\bibfnamefont {E.~K.}\ \bibnamefont
  {Kovev}}, \bibinfo {author} {\bibfnamefont {O.~N.}\ \bibnamefont {Efimov}},\
  and\ \bibinfo {author} {\bibfnamefont {L.~I.}\ \bibnamefont {Korovin}},\
  }\bibfield  {title} {\bibinfo {title} {Characteristics of anomalous
  transmission of x-rays in the general case of laue diffraction},\ }\href
  {https://doi.org/https://doi.org/10.1002/pssb.19690350147} {\bibfield
  {journal} {\bibinfo  {journal} {Physica Status Solidi (b)}\ }\textbf
  {\bibinfo {volume} {35}},\ \bibinfo {pages} {455} (\bibinfo {year}
  {1969})}\BibitemShut {NoStop}%
\bibitem [{\citenamefont {Liao}(2011)}]{Liao:2010ig}%
  \BibitemOpen
  \bibfield  {author} {\bibinfo {author} {\bibfnamefont {W.}~\bibnamefont
  {Liao}},\ }\bibfield  {title} {\bibinfo {title} {{Generation and search of
  axion-like light particle using intense crystalline field}},\ }\href
  {https://doi.org/10.1016/j.physletb.2011.06.064} {\bibfield  {journal}
  {\bibinfo  {journal} {Phys. Lett. B}\ }\textbf {\bibinfo {volume} {702}},\
  \bibinfo {pages} {55} (\bibinfo {year} {2011})},\ \Eprint
  {https://arxiv.org/abs/1011.6460} {arXiv:1011.6460 [hep-ph]} \BibitemShut
  {NoStop}%
\bibitem [{Note1()}]{Note1}%
  \BibitemOpen
  \bibinfo {note} {From
  https://x-server.gmca.aps.anl.gov/x0h.html.}\BibitemShut {Stop}%
\bibitem [{\citenamefont {Wark}\ and\ \citenamefont {Lee}(1999)}]{wark}%
  \BibitemOpen
  \bibfield  {author} {\bibinfo {author} {\bibfnamefont {J.}~\bibnamefont
  {Wark}}\ and\ \bibinfo {author} {\bibfnamefont {R.}~\bibnamefont {Lee}},\
  }\bibfield  {title} {\bibinfo {title} {{Simulations of femtosecond X-ray
  diffraction from unperturbed and rapidly heated single crystals}},\ }\href
  {https://doi.org/10.1107/s0021889899002861} {\bibfield  {journal} {\bibinfo
  {journal} {Journal of Applied Crystallography}\ }\textbf {\bibinfo {volume}
  {32}},\ \bibinfo {pages} {692} (\bibinfo {year} {1999})}\BibitemShut
  {NoStop}%
\bibitem [{\citenamefont {Shvyd'ko}\ and\ \citenamefont
  {Lindberg}(2012)}]{Shvydko:2012rzc}%
  \BibitemOpen
  \bibfield  {author} {\bibinfo {author} {\bibfnamefont {Y.}~\bibnamefont
  {Shvyd'ko}}\ and\ \bibinfo {author} {\bibfnamefont {R.}~\bibnamefont
  {Lindberg}},\ }\bibfield  {title} {\bibinfo {title} {{Spatiotemporal Response
  of Crystals in X-ray Bragg Diffraction}},\ }\href
  {https://doi.org/10.1103/PhysRevSTAB.15.100702} {\bibfield  {journal}
  {\bibinfo  {journal} {Phys. Rev. ST Accel. Beams}\ }\textbf {\bibinfo
  {volume} {15}},\ \bibinfo {pages} {100702} (\bibinfo {year} {2012})},\
  \Eprint {https://arxiv.org/abs/1207.3376} {arXiv:1207.3376 [physics.optics]}
  \BibitemShut {NoStop}%
\bibitem [{Note2()}]{Note2}%
  \BibitemOpen
  \bibinfo {note} {From
  https://x-server.gmca.aps.anl.gov/x0h.html.}\BibitemShut {Stop}%
\bibitem [{\citenamefont {Mozzanica}\ \emph {et~al.}(2016)\citenamefont
  {Mozzanica}, \citenamefont {Bergamaschi}, \citenamefont {Brueckner},
  \citenamefont {Cartier}, \citenamefont {Dinapoli}, \citenamefont
  {Greiffenberg}, \citenamefont {Jungmann-Smith}, \citenamefont {Maliakal},
  \citenamefont {Mezza}, \citenamefont {Ramilli}, \citenamefont {Ruder},
  \citenamefont {Schaedler}, \citenamefont {Schmitt}, \citenamefont {Shi},\
  and\ \citenamefont {Tinti}}]{Mozzanica2016}%
  \BibitemOpen
  \bibfield  {author} {\bibinfo {author} {\bibfnamefont {A.}~\bibnamefont
  {Mozzanica}}, \bibinfo {author} {\bibfnamefont {A.}~\bibnamefont
  {Bergamaschi}}, \bibinfo {author} {\bibfnamefont {M.}~\bibnamefont
  {Brueckner}}, \bibinfo {author} {\bibfnamefont {S.}~\bibnamefont {Cartier}},
  \bibinfo {author} {\bibfnamefont {R.}~\bibnamefont {Dinapoli}}, \bibinfo
  {author} {\bibfnamefont {D.}~\bibnamefont {Greiffenberg}}, \bibinfo {author}
  {\bibfnamefont {J.}~\bibnamefont {Jungmann-Smith}}, \bibinfo {author}
  {\bibfnamefont {D.}~\bibnamefont {Maliakal}}, \bibinfo {author}
  {\bibfnamefont {D.}~\bibnamefont {Mezza}}, \bibinfo {author} {\bibfnamefont
  {M.}~\bibnamefont {Ramilli}}, \bibinfo {author} {\bibfnamefont
  {C.}~\bibnamefont {Ruder}}, \bibinfo {author} {\bibfnamefont
  {L.}~\bibnamefont {Schaedler}}, \bibinfo {author} {\bibfnamefont
  {B.}~\bibnamefont {Schmitt}}, \bibinfo {author} {\bibfnamefont
  {X.}~\bibnamefont {Shi}},\ and\ \bibinfo {author} {\bibfnamefont
  {G.}~\bibnamefont {Tinti}},\ }\bibfield  {title} {\bibinfo {title}
  {Characterization results of the jungfrau full scale readout asic},\ }\href
  {https://doi.org/10.1088/1748-0221/11/02/C02047} {\bibfield  {journal}
  {\bibinfo  {journal} {Journal of Instrumentation}\ }\textbf {\bibinfo
  {volume} {11}}\bibinfo  {number} { (02)},\ \bibinfo {pages}
  {C02047}}\BibitemShut {NoStop}%
\bibitem [{\citenamefont {Maltezopoulos}\ \emph {et~al.}(2019)\citenamefont
  {Maltezopoulos}, \citenamefont {Dietrich}, \citenamefont {Freund},
  \citenamefont {Jastrow}, \citenamefont {Koch}, \citenamefont {Laksman},
  \citenamefont {Liu}, \citenamefont {Planas}, \citenamefont {Sorokin},
  \citenamefont {Tiedtke},\ and\ \citenamefont
  {Gr{\"{u}}nert}}]{Maltezopoulos2019}%
  \BibitemOpen
\bibfield  {number} {  }\bibfield  {author} {\bibinfo {author} {\bibfnamefont
  {T.}~\bibnamefont {Maltezopoulos}}, \bibinfo {author} {\bibfnamefont
  {F.}~\bibnamefont {Dietrich}}, \bibinfo {author} {\bibfnamefont
  {W.}~\bibnamefont {Freund}}, \bibinfo {author} {\bibfnamefont {U.~F.}\
  \bibnamefont {Jastrow}}, \bibinfo {author} {\bibfnamefont {A.}~\bibnamefont
  {Koch}}, \bibinfo {author} {\bibfnamefont {J.}~\bibnamefont {Laksman}},
  \bibinfo {author} {\bibfnamefont {J.}~\bibnamefont {Liu}}, \bibinfo {author}
  {\bibfnamefont {M.}~\bibnamefont {Planas}}, \bibinfo {author} {\bibfnamefont
  {A.~A.}\ \bibnamefont {Sorokin}}, \bibinfo {author} {\bibfnamefont
  {K.}~\bibnamefont {Tiedtke}},\ and\ \bibinfo {author} {\bibfnamefont
  {J.}~\bibnamefont {Gr{\"{u}}nert}},\ }\bibfield  {title} {\bibinfo {title}
  {{Operation of X-ray gas monitors at the European XFEL}},\ }\href
  {https://doi.org/10.1107/S1600577519003795} {\bibfield  {journal} {\bibinfo
  {journal} {Journal of Synchrotron Radiation}\ }\textbf {\bibinfo {volume}
  {26}},\ \bibinfo {pages} {1045} (\bibinfo {year} {2019})}\BibitemShut
  {NoStop}%
\bibitem [{\citenamefont {Emma}\ \emph {et~al.}(2017)\citenamefont {Emma},
  \citenamefont {Lutman}, \citenamefont {Guetg}, \citenamefont {Krzywinski},
  \citenamefont {Marinelli}, \citenamefont {Wu},\ and\ \citenamefont
  {Pellegrini}}]{Emma:2017qcq}%
  \BibitemOpen
  \bibfield  {author} {\bibinfo {author} {\bibfnamefont {C.}~\bibnamefont
  {Emma}}, \bibinfo {author} {\bibfnamefont {A.}~\bibnamefont {Lutman}},
  \bibinfo {author} {\bibfnamefont {M.~W.}\ \bibnamefont {Guetg}}, \bibinfo
  {author} {\bibfnamefont {J.}~\bibnamefont {Krzywinski}}, \bibinfo {author}
  {\bibfnamefont {A.}~\bibnamefont {Marinelli}}, \bibinfo {author}
  {\bibfnamefont {J.}~\bibnamefont {Wu}},\ and\ \bibinfo {author}
  {\bibfnamefont {C.}~\bibnamefont {Pellegrini}},\ }\bibfield  {title}
  {\bibinfo {title} {{Experimental demonstration of fresh bunch self-seeding in
  an X-ray free electron laser}},\ }\href {https://doi.org/10.1063/1.4980092}
  {\bibfield  {journal} {\bibinfo  {journal} {Appl. Phys. Lett.}\ }\textbf
  {\bibinfo {volume} {110}},\ \bibinfo {pages} {154101} (\bibinfo {year}
  {2017})}\BibitemShut {NoStop}%
\bibitem [{\citenamefont {Junk}(1999)}]{Junk:1999kv}%
  \BibitemOpen
  \bibfield  {author} {\bibinfo {author} {\bibfnamefont {T.}~\bibnamefont
  {Junk}},\ }\bibfield  {title} {\bibinfo {title} {{Confidence level
  computation for combining searches with small statistics}},\ }\href
  {https://doi.org/10.1016/S0168-9002(99)00498-2} {\bibfield  {journal}
  {\bibinfo  {journal} {Nucl. Instrum. Meth. A}\ }\textbf {\bibinfo {volume}
  {434}},\ \bibinfo {pages} {435} (\bibinfo {year} {1999})},\ \Eprint
  {https://arxiv.org/abs/hep-ex/9902006} {arXiv:hep-ex/9902006} \BibitemShut
  {NoStop}%
\bibitem [{\citenamefont {Della~Valle}\ \emph {et~al.}(2014)\citenamefont
  {Della~Valle}, \citenamefont {Ejlli}, \citenamefont {Gastaldi}, \citenamefont
  {Messineo}, \citenamefont {Milotti}, \citenamefont {Pengo}, \citenamefont
  {Piemontese}, \citenamefont {Ruoso},\ and\ \citenamefont
  {Zavattini}}]{DellaValle:2014wea}%
  \BibitemOpen
  \bibfield  {author} {\bibinfo {author} {\bibfnamefont {F.}~\bibnamefont
  {Della~Valle}}, \bibinfo {author} {\bibfnamefont {A.}~\bibnamefont {Ejlli}},
  \bibinfo {author} {\bibfnamefont {U.}~\bibnamefont {Gastaldi}}, \bibinfo
  {author} {\bibfnamefont {G.}~\bibnamefont {Messineo}}, \bibinfo {author}
  {\bibfnamefont {E.}~\bibnamefont {Milotti}}, \bibinfo {author} {\bibfnamefont
  {R.}~\bibnamefont {Pengo}}, \bibinfo {author} {\bibfnamefont
  {L.}~\bibnamefont {Piemontese}}, \bibinfo {author} {\bibfnamefont
  {G.}~\bibnamefont {Ruoso}},\ and\ \bibinfo {author} {\bibfnamefont
  {G.}~\bibnamefont {Zavattini}},\ }\bibfield  {title} {\bibinfo {title} {{New
  PVLAS model independent limit for the axion coupling to $\gamma\gamma$ for
  axion masses above 1 meV}},\ }in\ \href
  {https://doi.org/10.3204/DESY-PROC-2014-03/gastaldi_ugo} {\emph {\bibinfo
  {booktitle} {{10th Patras Workshop on Axions, WIMPs and WISPs}}}}\ (\bibinfo
  {year} {2014})\ pp.\ \bibinfo {pages} {67--70},\ \Eprint
  {https://arxiv.org/abs/1410.4081} {arXiv:1410.4081 [hep-ex]} \BibitemShut
  {NoStop}%
\bibitem [{\citenamefont {Battesti}\ \emph {et~al.}(2010)\citenamefont
  {Battesti}, \citenamefont {Fouche}, \citenamefont {Detlefs}, \citenamefont
  {Roth}, \citenamefont {Berceau}, \citenamefont {Duc}, \citenamefont {Frings},
  \citenamefont {Rikken},\ and\ \citenamefont {Rizzo}}]{Battesti:2010dm}%
  \BibitemOpen
  \bibfield  {author} {\bibinfo {author} {\bibfnamefont {R.}~\bibnamefont
  {Battesti}}, \bibinfo {author} {\bibfnamefont {M.}~\bibnamefont {Fouche}},
  \bibinfo {author} {\bibfnamefont {C.}~\bibnamefont {Detlefs}}, \bibinfo
  {author} {\bibfnamefont {T.}~\bibnamefont {Roth}}, \bibinfo {author}
  {\bibfnamefont {P.}~\bibnamefont {Berceau}}, \bibinfo {author} {\bibfnamefont
  {F.}~\bibnamefont {Duc}}, \bibinfo {author} {\bibfnamefont {P.}~\bibnamefont
  {Frings}}, \bibinfo {author} {\bibfnamefont {G.~L. J.~A.}\ \bibnamefont
  {Rikken}},\ and\ \bibinfo {author} {\bibfnamefont {C.}~\bibnamefont
  {Rizzo}},\ }\bibfield  {title} {\bibinfo {title} {{A Photon Regeneration
  Experiment for Axionlike Particle Search using X-rays}},\ }\href
  {https://doi.org/10.1103/PhysRevLett.105.250405} {\bibfield  {journal}
  {\bibinfo  {journal} {Phys. Rev. Lett.}\ }\textbf {\bibinfo {volume} {105}},\
  \bibinfo {pages} {250405} (\bibinfo {year} {2010})},\ \Eprint
  {https://arxiv.org/abs/1008.2672} {arXiv:1008.2672 [hep-ex]} \BibitemShut
  {NoStop}%
\bibitem [{\citenamefont {Inada}\ \emph {et~al.}(2017)\citenamefont {Inada}
  \emph {et~al.}}]{Inada:2016jzh}%
  \BibitemOpen
  \bibfield  {author} {\bibinfo {author} {\bibfnamefont {T.}~\bibnamefont
  {Inada}} \emph {et~al.},\ }\bibfield  {title} {\bibinfo {title} {{Search for
  Two-Photon Interaction with Axionlike Particles Using High-Repetition Pulsed
  Magnets and Synchrotron X Rays}},\ }\href
  {https://doi.org/10.1103/PhysRevLett.118.071803} {\bibfield  {journal}
  {\bibinfo  {journal} {Phys. Rev. Lett.}\ }\textbf {\bibinfo {volume} {118}},\
  \bibinfo {pages} {071803} (\bibinfo {year} {2017})},\ \Eprint
  {https://arxiv.org/abs/1609.05425} {arXiv:1609.05425 [hep-ex]} \BibitemShut
  {NoStop}%
\bibitem [{\citenamefont {O'Hare}(2024)}]{OHare:2024nmr}%
  \BibitemOpen
  \bibfield  {author} {\bibinfo {author} {\bibfnamefont {C.~A.~J.}\
  \bibnamefont {O'Hare}},\ }\bibfield  {title} {\bibinfo {title} {{Cosmology of
  axion dark matter}},\ }\href {https://doi.org/10.22323/1.454.0040} {\bibfield
   {journal} {\bibinfo  {journal} {PoS}\ }\textbf {\bibinfo {volume}
  {COSMICWISPers}},\ \bibinfo {pages} {040} (\bibinfo {year} {2024})},\ \Eprint
  {https://arxiv.org/abs/2403.17697} {arXiv:2403.17697 [hep-ph]} \BibitemShut
  {NoStop}%
\bibitem [{\citenamefont {Rabadan}\ \emph {et~al.}(2006)\citenamefont
  {Rabadan}, \citenamefont {Ringwald},\ and\ \citenamefont
  {Sigurdson}}]{Rabadan:2005dm}%
  \BibitemOpen
  \bibfield  {author} {\bibinfo {author} {\bibfnamefont {R.}~\bibnamefont
  {Rabadan}}, \bibinfo {author} {\bibfnamefont {A.}~\bibnamefont {Ringwald}},\
  and\ \bibinfo {author} {\bibfnamefont {K.}~\bibnamefont {Sigurdson}},\
  }\bibfield  {title} {\bibinfo {title} {{Photon regeneration from
  pseudoscalars at X-ray laser facilities}},\ }\href
  {https://doi.org/10.1103/PhysRevLett.96.110407} {\bibfield  {journal}
  {\bibinfo  {journal} {Phys. Rev. Lett.}\ }\textbf {\bibinfo {volume} {96}},\
  \bibinfo {pages} {110407} (\bibinfo {year} {2006})},\ \Eprint
  {https://arxiv.org/abs/hep-ph/0511103} {arXiv:hep-ph/0511103} \BibitemShut
  {NoStop}%
\end{thebibliography}%
\end{document}